\pdfoutput=1

\documentclass[aps,twocolumn,amsmath,amssymb,preprintnumbers,floatfix,prb,superscriptaddress,longbibliography]{revtex4-2}%{revtex4}%

\usepackage[utf8]{inputenc}
\usepackage{newtxtext}
\usepackage[upint]{newtxmath}
\usepackage{microtype}
\usepackage{textcomp}
\usepackage{eucal}
\usepackage{bm}
\usepackage{siunitx}
\usepackage{comment}
\usepackage{lipsum}

\usepackage{enumerate}
\usepackage{amsfonts}
\usepackage{amsmath}
\usepackage{amssymb}
\usepackage{color}
\usepackage{soul}

\usepackage{graphicx}

\usepackage[colorlinks,allcolors=blue]{hyperref}
\usepackage[capitalize]{cleveref}

\definecolor{DarkRed}{rgb}{0.65,0,0}%
\definecolor{Green}{rgb}{0,0.3,0.3}
\definecolor{Purple}{rgb}{0.3,0,0.65}
\definecolor{Red}{rgb}{1,0,0}
\definecolor{Blue}{rgb}{0,0,0.85}
\definecolor{Magenta}{rgb}{1,0,1}

\newcommand{\Imag}{{\Im\mathrm{m}}}   % Imaginary part 
\newcommand{\Real}{{\mathrm{Re}}}   % Real part
\newcommand{\im}{\mathrm{i}}        % Imaginary unit non-italic
\newcommand{\ve}[1]{\boldsymbol{#1}}
\newcommand{\abs}[1]{|#1|}

\newcommand{\veck}{\ve{k}}

\def\i{\mathrm{i}}

\newcommand{\g}{\underline{\gamma}}

\newcommand{\be}{\begin{equation}}
\newcommand{\ee}{\end{equation}}

\newcommand{\prlsection}[1]{\textit{#1}.\kern0.05em---\kern0.05em\ignorespaces}

\begin{document}
\title{Spin pumping from a ferromagnetic insulator into an altermagnet}
\author{Chi Sun}
\affiliation{Center for Quantum Spintronics, Department of Physics, Norwegian \\ University of Science and Technology, NO-7491 Trondheim, Norway}
\author{Jacob Linder}
%\email[Corresponding author: ]{jacob.linder@ntnu.no}
\affiliation{Center for Quantum Spintronics, Department of Physics, Norwegian \\ University of Science and Technology, NO-7491 Trondheim, Norway}

\begin{abstract}
A class of antiferromagnets with spin-polarized electron bands, yet zero net magnetization, called altermagnets is attracting increasing attention due to their potential use in spintronics. Here, we study spin injection into an altermagnet via spin pumping from a ferromagnetic insulator. We find that the spin pumping behaves qualitatively different depending on how the altermagnet is crystallographically oriented relative the interface to the ferromagnetic insulator. The altermagnetic state can enhance or suppress spin pumping, which we explain in terms of spin-split altermagnetic band structure and the spin-flip probability for the incident modes. Including the effect of interfacial Rashba spin-orbit coupling, we find that the spin-pumping effect is in general magnified, but that it can display a non-monotonic behavior as a function of the spin-orbit coupling strength. We show that there exists an optimal value of the spin-orbit coupling strength which causes an order of magnitude increase in the pumped spin current, even for the crystallographic orientation of the altermagnet which suppresses the spin pumping.
\end{abstract}
\maketitle

\textit{Introduction. --} Spin pumping is a mechanism for generating spin currents in which the precessing magnetization in a magnetic material transfers angular momentum into its adjacent nonmagnetic layers \cite{tserkovnyak_rmp_04,Tserkovnyak2002Dec,Tserkovnyak2002Feb,Brataas2020Nov,Sun2022Mar}. Compared with metals, magnetic insulators can function as efficient spin-current sources with low dissipation and reduced energy loss \cite{Brataas2020Nov}, in which the ferromagnetic insulator (FI) YIG demonstrates the lowest known spin dissipation with an exceptionally low Gilbert damping \cite{Heinrich2011Aug,Haertinger2015Aug}. In conventional FI/normal metal (NM) heterostructures, the injected spin current affects the magnetization dynamics in the FI and creates a spin accumulation in the NM, resulting in a measurable damping increase in the linewidth of a ferromagnetic resonance (FMR) signal, which has been extensively investigated \cite{fmr_sp,Rezende2013enhanced,Tserkovnyak2002Feb,Castel2012Oct}. When the NM is replaced by another material such as a superconductor, the spin pumping effect is considerably modulated by various superconducting gap properties and interfacial effects \cite{Sun2023Apr,Jeon2018Jun,Yao2018Jun,Carreira2021Oct,inoue_17,kato_prb_2019,ominato_prb_2022,ominato_prb_2022a}.

Recently, a new magnetic phase dubbed altermagnetism \cite{ahn_prb_2019, hayami_jpsj_19, smejkal_sa_20, yuan_prb_20} has attracted increasing attention. Such materials exhibit a large momentum-dependent spin-splitting and vanishing net macroscopic magnetization at the same time, 
thus combining features from conventional ferromagnets and antiferromagnets \cite{am_emerging_22,vsmejkal_2022beyond,mazin2022_altermagnetism,mazin2023_altermagnetism}. The spin splitting in the altermagnet (AM), which is of a strong non relativistic origin, is protected by the broken symmetries of the spin arrangements on the crystal, distinct from ferromagnetic and relativistically spin-orbit coupled (SOC) systems \cite{am_emerging_22,vsmejkal_2022beyond,SP_SOC_Shufeng}. It is predicted that AM can span a large range of materials, from insulators like FeF$_2$ and MnF$_2$, semiconductors like MnTe, metals like RuO$_2$, to superconductors like La$_2$CuO$_4$ \cite{am_emerging_22,RuO2_22,RuO2_obser_23,Mnte_23}. These novel properties make AM a fascinating material platform to investigate superconducting \cite{Sun2023_andreev,ouassou_prl_23,mazin2022_altermagnetism, zhang_arxiv_2023, Papaj2023May, beenakker_prb_23} and spintronics phenomena 
\cite{magnon_magnon_AM,spin_space_group_AM1, das_arxiv_23, spin_space_group_AM2,bai_prl23}.

\begin{figure}[t!]
\includegraphics[width=0.65\columnwidth]{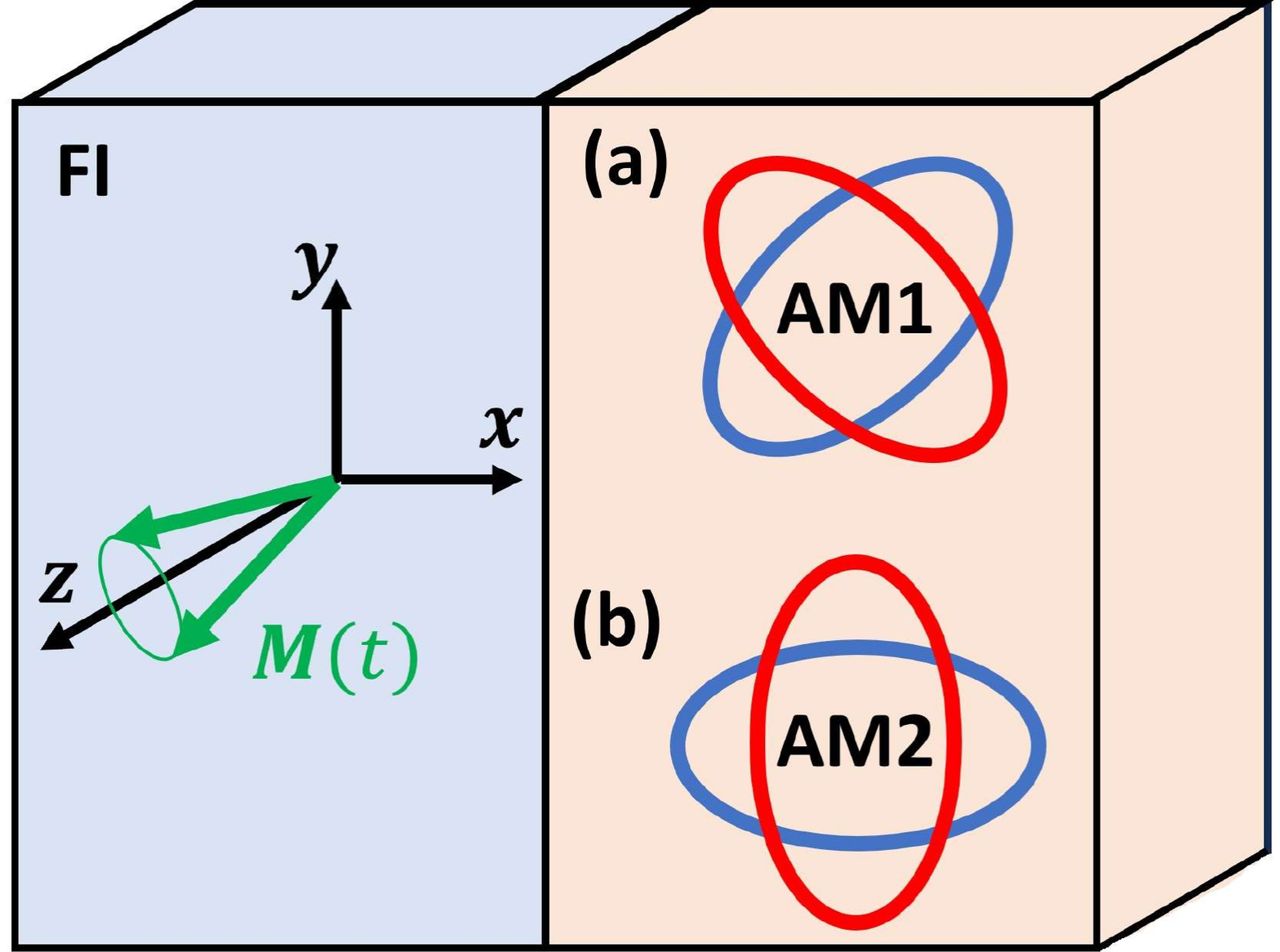}
	\caption{(Color online) Spin pumping is considered in a bilayer consisting of a ferromagnetic insulator (FI) and an altermagnet (AM). The magnetization $\boldsymbol{M}(t)$ in the FI is precessing around the $z$ axis at the FMR frenquency $\Omega$. Different interface orientations are also considered, effectively rotating the spin-resolved Fermi surface in the AM for $e\uparrow$ (red ellipse) and $e\downarrow$ (blue ellipse) spin carriers. For notation simplicity, the two AM orientations are referred as AM1 and AM2, respectively.}
	\label{fig:structure}
\end{figure}

In this work, we theoretically determine spin pumping from a FI into a metallic AM in a FI/AM bilayer (see Fig. \ref{fig:structure}). To cover different crystallographic orientations of the interface relative to the spin-polarized lobes of the altermagnetic Fermi surface, two representative metallic AMs, as shown in Figs. \ref{fig:structure}(a) and \ref{fig:structure}(b), are studied in detail. In addition to the non relativistic interfacial effect induced by the AM, a relativistic Rashba SOC is included at the FI/AM interface in our model. We find that the spin pumping current can be  enhanced or suppressed by altermagnetism, depending on the interface orientation, thus offering versatility. This is explained in terms of the spin-split altermagnetic band structure and the spin-flip probability for the incident modes toward the interface. In addition, the spin pumping current shows a non-monotonic behavior as a function of the interfacial SOC strength. We show that the interfacial SOC can, in a certain range, increase the spin pumping current in a FI/AM bilayer by more than an order of magnitude. 

\textit{Theory. --} The effective low-energy Hamiltonian for the AM shown in Fig. \ref{fig:structure}(a), using an electron field operator basis $\psi = [\psi_\uparrow, \psi_\downarrow]^T$, is given by
\begin{equation}
H_\text{AM} =  -{\frac{\hbar^2\triangledown^2}{2 m_e}} - \mu + \alpha\sigma_zk_x k_y,
\end{equation}
in which $\alpha$ is the parameter characterizes the altermagnetism strength, $\sigma_z$ denotes the Pauli matrix, $m_e$ is the electron mass and $\mu$ is the chemical potential. By solving the stationary Schr\"odinger equation as an eigenvalue problem (see SM for details), the $x$-components of the wave vectors in the AM with energy $E$ are given by $k_{e\uparrow(\downarrow),\pm}=\pm{\hbar}^{-1}\sqrt{2m_e(\mu+E)-\hbar^2 k_y^2 +{\alpha^2 m_e^2 k_y^2}/{\hbar^2}}\mp^{'}{\alpha m_e k_y}/{\hbar^2}$, in which the $\pm$ sign denotes the propagation direction along the $\pm x$, $e\uparrow(\downarrow)$ describes electron with spin up (down), and $\mp^{'}=-(+)$ for $\uparrow(\downarrow)$. Here we assume translational invariance in the $y$-direction with belonging momentum $k_y$ of the incident particle.

On the other hand, the Hamiltonian for the FI has the form
\begin{equation}
    H_{\text{FI}} =  -{\frac{\hbar^2\triangledown^2}{2 m_e}} + U + J\hat{\boldsymbol{\sigma}}\cdot \boldsymbol{M}(t),
\end{equation}
in which $\hat{\boldsymbol{\sigma}}$ denotes the Pauli matrix vector and $J$ is the exchange interaction. Here the potential $U$ is larger than $\mu$ in the nearby AM to ensure the ferromagnet to be insulating. The normalized magnetization is defined as $\boldsymbol{M}(t) = (m \cos\Omega t, m \sin \Omega t, \sqrt{1-m^2})$, where $m \in [0,1]$ is the magnetization oscillation amplitude and $\Omega$ denotes the FMR frequency for spin pumping. By employing a wavefunction with the structure $(e^{-\frac{i \Omega t}{2}},e^{\frac{i \Omega t}{2}})^T$ for its additional time-dependence, the non-stationary Schr\"odinger equation can be solved as an eigenvalue problem (see SM for details). The two eigenpairs are obtained as: $E_1=E_+$ with $(a_+,b_+)^T$ and $E_2=E_-$ with $(a_-,b_-)^T$, in which $E_\pm=U+\frac{\hbar^2(k_{x}^2+k_{y}^2)}{2m_e}\pm JR$ with $R=(1-2\beta\sqrt{1-m^2}+\beta^2)^{1/2}$ and $\beta={\hbar\Omega}/2J$.

To study the spin pumping effect, we first consider an $e\uparrow$ incident electron with excitation energy $E$ from the AM side based on the FI/AM bilayer. The wavefunctions are given by
\begin{align}
\Psi_{\text{AM},e\uparrow} &=
\left[\begin{pmatrix}
1\\0
\end{pmatrix}e^{ik_{e\uparrow,-}x}+r
\begin{pmatrix}
 1\\0   
\end{pmatrix}e^{ik_{e\uparrow,+}x}\right]e^{-\frac{iEt}{\hbar}} \notag\\
&+ r^{'}
\begin{pmatrix}
    0\\1
\end{pmatrix}e^{ik^{'}_{e\downarrow,+}x}e^{-\frac{iE^{'}t}{\hbar}}, 
\label{eq:AM_e_up_main}\\
\Psi_{\text{FI},e\uparrow} &=t
\begin{pmatrix}
 a_+e^{\frac{-i\Omega t}{2}}\\b_+e^{\frac{i\Omega t}{2}}  
\end{pmatrix}e^{-ik_{\text{F}1,e\uparrow}x}e^{\frac{-iE_1t}{\hbar}} \notag\\
&+p\begin{pmatrix}
    a_-e^{\frac{-i\Omega t}{2}}\\b_-e^{\frac{i\Omega t}{2}}
\end{pmatrix}e^{-ik_{\text{F}2,e\uparrow}x}e^{\frac{-iE_2t}{\hbar}},
\label{eq:FI_e_up_main}
\end{align}
in which $r$ and $r^{'}$ are coefficients describing reflection without and with spin-flip in the AM, respectively, and $t$ and $p$ are transmission coefficients in the FI. To differentiate it from the incident energy $E$, the energy after the spin-flip in the AM due to spin pumping is denoted as $E^{'}$. By matching the time-dependence of the wavefunction components on the AM and FI sides, we obtain $E^{'}=E-\hbar\Omega$ and $E_1=E_2=E-\frac{\hbar\Omega}{2}$. In terms of $E$, the corresponding $x$-component of the two wave vectors in the FI are expressed as 
$k_{\text{F}1,e\uparrow}=\hbar^{-1}\sqrt{2m_e[E-U-J(R+\beta)]-\hbar^2k_y^2}$ and $k_{\text{F}2,e\uparrow}=\hbar^{-1}\sqrt{2m_e[E-U+J(R-\beta)]-\hbar^2k_y^2}$. Note that the wave numbers in the FI possess imaginary values due to a large potential $U$, ensuring evanescent electron states in the FI. Details of the wave functions induced by an $e\downarrow$ incident partice with excitation energy $E$ from the AM can be found in the SM, in which we have $E^{'}=E+\hbar\Omega$.

Appropriate boundary conditions are required to solve the reflection and transmissions coefficients in the wavefunctions. Here we consider a Rashba spin-orbit coupled interface with the Hamiltonian
\begin{equation}
H_I = [U_0 + \frac{U_\text{SO}}{k_F} \hat{\boldsymbol{x}} \cdot (\hat{\boldsymbol{\sigma}} \times \boldsymbol{k})]\delta(x)=[U_0 - \frac{U_\text{SO}}{k_F}k_y \sigma_z]\delta(x),
\end{equation}
in which $U_0$ is the interfacial energy barrier, $U_\text{SO}$ describes the Rashba SOC, $k_F=\sqrt{2m_e\mu}/\hbar$ is the Fermi wave vector and $\hat{\boldsymbol{x}}$ denotes the interface normal. On the other hand, to derive the boundary condition, antisymmetrization of the altermagnetic term $\alpha k_xk_y\sigma_z \to \frac{\alpha k_y}{2} \{k_x,\Theta(x)\} {\sigma_z}$ is necessary to ensure hermiticity of the Hamilton-operator, where $\Theta(x)$ is the step function and $k_x = -\i\partial_x$. Combing all related Hamiltonian contributions in the FI/AM system, we obtain $\Psi_{\text{AM},e\uparrow}\big|_{x=0}=\Psi_{\text{FI},e\uparrow}\big|_{x=0}=(f,g)^T$ and
\begin{align}
\partial_x\Psi_{\text{AM},e\uparrow}\big|_{x=0}-\partial_x\Psi_{\text{FI},e\uparrow}\big|_{x=0}=\begin{pmatrix}
k_{\alpha,+1}f\\k_{\alpha,-1}g
\end{pmatrix},
\label{eq:BC_main}
\end{align} 
where $k_{\alpha,\sigma} =\frac{2m_e}{\hbar^2}[U_0-(\frac{\i\alpha}{2}+\frac{U_\text{SO}}{k_F})k_y\sigma]$ with $\sigma=+1(-1)$. Here the imaginary number $\im$ appears in $k_{\alpha,\sigma}$ since we consider $k_y$ invariance (unlike $k_x = -\im\partial_x$). Note that the boundary conditions for $e\downarrow$ incident from the AM side have the same forms as $e\uparrow$ with different explicit expressions of $f$ and $g$ in the wave functions. 

The longitudinal quantum mechanical spin current polarized along the $z$ axis in the AM is given by 
\begin{equation}
j_{sz,e\uparrow(\downarrow)}=\frac{\hbar^2}{2m_e}(\Imag\{f^*\nabla f\} - \Imag\{g^*\nabla g\} )+\frac{\alpha k_y}{2}(|f|^2+|g|^2).
\end{equation}
Integrating over all energies and all possible transverse modes via $\int dk_x = \int dE (dk_x/dE)$ and $\int dk_y$, the spin pumping current is calculated as
\begin{equation}
I_{s,e\uparrow(\downarrow)}=\int dk_y \int dE \frac{dk_x}{dE} j_{sz,e\uparrow(\downarrow)}f_0(E),
\end{equation}
in which $f_0(E)$ denotes the Fermi-Dirac distribution. Note that $dk_x/dE$ plays the role of 1D DOS in the AM instead of 2D DOS since here $\int dk_y$ is included separately. Including contributions from both $e\uparrow$ and $e\downarrow$ incidents, the total spin pumping current is $I_s = I_{s,e\uparrow} + I_{s,e\downarrow}$. In general, a backflow spin current exists due to a spin accumulation that is built up in the material connected to the precessessing FI \cite{tserkovnyak_rmp_04}, which diminishes the magnitude of the total spin current flowing across the interface. The backflow spin current can safely be neglected in the present case of a ballistic large AM reservoir.
%In general, a backflow spin current exists due to a spin accumulation that is built up in the material connected to the precessessing FI \cite{tserkovnyak_rmp_04}. This backflow current diminishes the magnitude of the total spin current flowing across the interface. Assuming that the material which the spin current is pumped into act as a highly conductive reservoir which drains the spin current, the backflow spin current may be neglected. For a ferromagnet/normal metal bilayer with a Rashba spin-orbit coupled interface, as in the present system, Ref. \cite{chen_prl_15} derived a backflow factor $\xi \propto (\lambda_\text{sd}/l_\text{mfp}) \text{coth}(d_N/\lambda_\text{sd})$ where $\lambda_\text{sd}$ is the spin diffusion length, $l_\text{mfp}$ is the electronic mean free path, and $d_N$ is the thickness of the normal layer. For ballistic, large reservoirs, $\xi \to 0$.
To show how the crystallographic orientation of the interface between the materials affects the spin pumping, the AM corresponding to a $45$ degree rotation of the interface, as shown in Fig. \ref{fig:structure}(b), is modeled by replacing $\alpha k_xk_y \to \alpha(k_x^2-k_y^2)/2$ in $H_\text{AM}$. This leads to different expressions for the wavevectors, boundary conditions and quantum mechanical spin pumping current (see SM for details). Our model can also be expanded to a AM with arbitrary rotation by combination of the established 0 and 45 degree cases, i.e., using $\alpha_1 k_x k_y{\sigma_z} + \alpha_2 (k_x^2- k_y^2){\sigma_z}/2$ in $H_\text{AM}$ with the arbitrary angle determined by $\theta_\alpha = \frac{1}{2}\arctan(\alpha_1/\alpha_2)$.

\textit{Results: Altermagnetism dependence. --} For notation simplicity, we refer the altermagnetic Fermi surface structures shown in Figs. \ref{fig:structure}(a) and \ref{fig:structure}(b) as AM1 and AM2, respectively, corresponding to different interface orientations by effectively rotating 45 degree of the spin-resolved Fermi-surfaces. To ensure each spin-polarized lobe of the altermagnetic Fermi surface described by $H_\text{AM}$ defines a closed integral path or ellipse rather than a hyperbola, $\alpha < \hbar^2/m_e \equiv \alpha_c$ should be satisfied (see SM for details). The semi-major (minor) axis $a$ ($b$) of the ellipse can be obtained as 
\begin{align}
a = \sqrt{\frac{2m_e(\mu+E)}{\hbar^2-m_e \alpha}},\; b = \sqrt{\frac{2m_e(\mu+E)}{\hbar^2+m_e\alpha}},
\end{align}
based on which $a$ ($b$) increases (decreases) with $\alpha$.

\begin{figure}[t!]
\includegraphics[width=0.99\columnwidth]{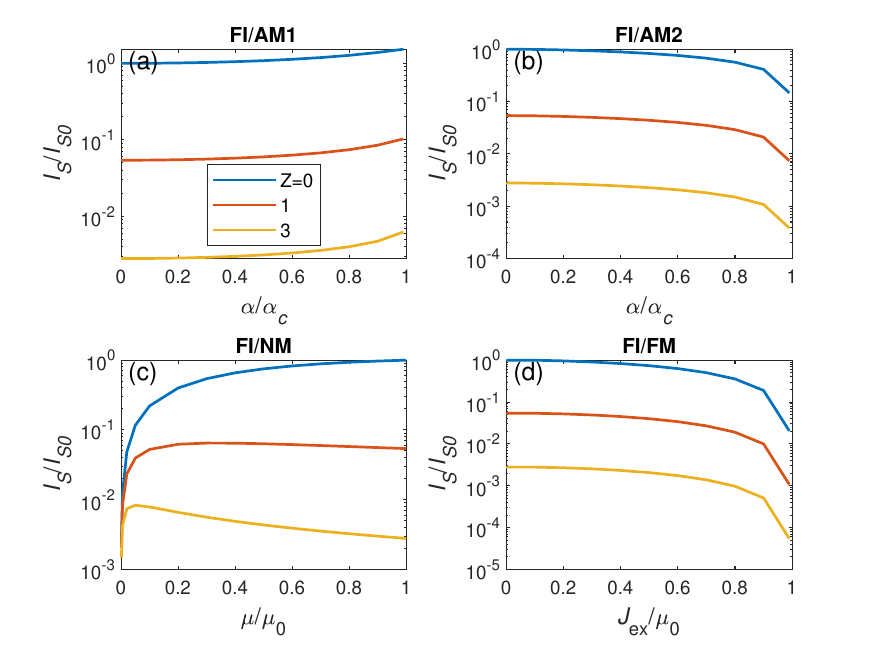}
	\caption{(Color online) Normalized spin pumping current $I_s/I_{s0}$ as a function of altermganetism for FI/AM1 and FI/AM2 in (a) and (b), respectively. (c) $I_s/I_{s0}$ as a function of chemical potential $\mu$ for FI/NM. (d)  $I_s/I_{s0}$ as a function of exchange energy $J_\text{ex}$ for FI/FM. In the absence of Rashba SOC, different interfacial barriers $Z=0,1,3$ are considered. Here $m=0.2$ and $\hbar\Omega=0.5$ meV are utilized. $I_{s0}$ corresponds to the spin pumping current for FI/NM with $\mu/\mu_0=1$.
	}
	\label{fig:alpha}
\end{figure}

In the absence of Rashba SOC, the dimensionless parameter $Z= \frac{m_e U_0}{\hbar^2 k_F}$ characterizes the quality of electric contact between the FI and AM. To model high-transparent to tunneling interfaces, we investigate the spin pumping current $I_s$ with $Z=0,1,3$ in Fig. \ref{fig:alpha}. As is reasonable, $I_s$ decreases as $Z$ increases. More importantly, we find that $I_s$ increases with $\alpha$ in FI/AM1 [Fig. \ref{fig:alpha}(a)] while it decreases with $\alpha$ in FI/AM2 [Fig. \ref{fig:alpha}(b)], indicating the crucial role of the interface orientation in FI/AM for spin pumping.

To understand the altermagnetism dependence behavior, it is instructional to consider the altermagnetic Fermi surfaces and energy bands. For simplicity, let us focus on particles close to normal incidence, $k_y \to 0$, which contribute the most to the transport across the junction. In AM1, the wavevectors of the $e\uparrow$ and $e\downarrow$ incident particles are the same, i.e., $k_{e\uparrow(\downarrow),\pm}=\pm{\hbar}^{-1}\sqrt{2m_e(\mu+E)}$, just like the NM case. This analogy also applies when integrating over all possible $k_y$ values, i.e., the total spin polarization of the incident particles cancels since spin-$\downarrow$ is the majority carrier for $k_y>0$ and spin-$\uparrow$ is the majority carrier for $k_y<0$ and the two spin bands contribute equally. On the other hand, in AM2, the wavevectors can be strongly mismatched even for $k_y\to 0$, i.e., $k_{e\uparrow,\pm}=\pm{\hbar}^{-1}\sqrt{2m_e(\mu+E)/(\hbar^2+m_e\alpha)}$ and $k_{e\downarrow,\pm}=\pm{\hbar}^{-1}\sqrt{2m_e(\mu+E)/(\hbar^2-m_e\alpha)}$. This is similar to the ferromagnetic metal (FM) case, in which a large mismatch between these wavevectors is induced by a (momentum-independent) spin-splitting or exchange
energy $J_\text{ex}$ by considering the Hamiltonian $H_\text{FM} =  -{\frac{\hbar^2\triangledown^2}{2 m_e}} - \mu + J_\text{ex}{\sigma_z}$. Therefore, it is useful to compare the spin pumping current based on FI/NM and FI/FM, as shown in Figs. \ref{fig:alpha}(c) and \ref{fig:alpha}(d), respectively.

The total spin current is determined by the spin-flip probability between $e\uparrow$ and $e\downarrow$ states induced by spin pumping, and also the number of available $k_y$ modes for spin-flip. Let us first consider the altermagnetism dependence of the number of $k_y$ modes. As discussed before, $a$ ($b$) increases (decreases) with $\alpha$. In AM1, the allowed number of $k_y$ mode or $\abs{k_y}$ maximum for both $e\uparrow$ and $e\downarrow$ bands increases with $\alpha$ as the semi-major axis $a$ increases, giving rise to more available transverse $k_y$ modes in which the the spin-flip between $e\uparrow$ and $e\downarrow$ can be realized. Note that the asymmetry between incident spin $e\uparrow$ and $e\downarrow$ is broken by the spin pumping FMR frequecy $\Omega$. Therefore, the total spin current $I_s$, which includes contributions from both $e\uparrow$ and $e\downarrow$ incidents, is enhanced when integrating over $k_y$. This is consistent with the trends shown in Fig. \ref{fig:alpha}(a). Similarly, the allowed $k_y$ range for spin-flip can be increased by increasing $\mu$ in the NM, giving rise to an enhanced $I_s$ with a high-transparent $Z=0$ interface [see blue curve in Fig. \ref{fig:alpha}(c)]. However, it can be seen that the trends change for large $Z$, indicating a difference between increasing $\alpha$ and $\mu$, although in both cases the number of $k_y$ states that carry spin current increases. This can be explained by considering the spin-flip probability for each $k_y$ mode, which we will get back to. 

On the other hand, in AM2, the allowed $k_y$ modes increase with increasing $\alpha$ and semi-major axis $a$ for the $e\uparrow$ band while they decrease with increasing $\alpha$ and decreasing semi-minor axis $b$ for the $e\downarrow$ band. This results in an enhanced mismatch between the spin-bands at a given value of $k_y$, and therefore less transverse modes available to realize spin-flip between the two bands. This corresponds to the trend that $I_s$ is suppressed with $\alpha$, as shown in Fig. \ref{fig:alpha}(b). The same mechanism applies for FM in Fig. \ref{fig:alpha}(d), in which the mismatch between available $k_y$ modes for $e\uparrow$ and $e\downarrow$ bands is enhanced with increasing $J_\text{ex}$, confirming the similarity between AM2 and FM.

Next, we turn to the spin-flip probability at a fixed $k_y$, in particular small $\abs{k_y}$ close to normal incidence which contribute the most. As calculated in detail in the SM (see Fig. \ref{fig:prob}), it is found that the spin-flip probability increases (decreases) with altermagnetism for FI/AM1(AM2) , which corresponds to the trends shown in Fig. \ref{fig:alpha}. The spin-flip probability behavior can be understood by considering the magnitude of momentum transfer (along $x$), e.g., when a (spin-flip) reflection requires a large momentum transfer, its probability is diminished \cite{momentum_transfer_r1,momentum_transfer_2}.
In AM1 (AM2), the magnitude of the momentum transfer [e.g., between $k_{e\uparrow,-}$ and $k_{e\downarrow,+}^{'}$ in Eq. (\ref{eq:AM_e_up_main})] at fixed $k_y$ decreases (increases) with altermagnetism. Similarly, in FI/NM, the magnitude of momentum transfer for spin-flip increases as $\mu$, which suppresses the spin-flip probability. This compensates the fact that more $k_y$ modes are available when $\mu$ increases, as discussed before, giving a total suppression of spin current for large $Z$ in Fig. \ref{fig:alpha}(c).

\begin{figure}[t!]
\includegraphics[width=0.99\columnwidth]{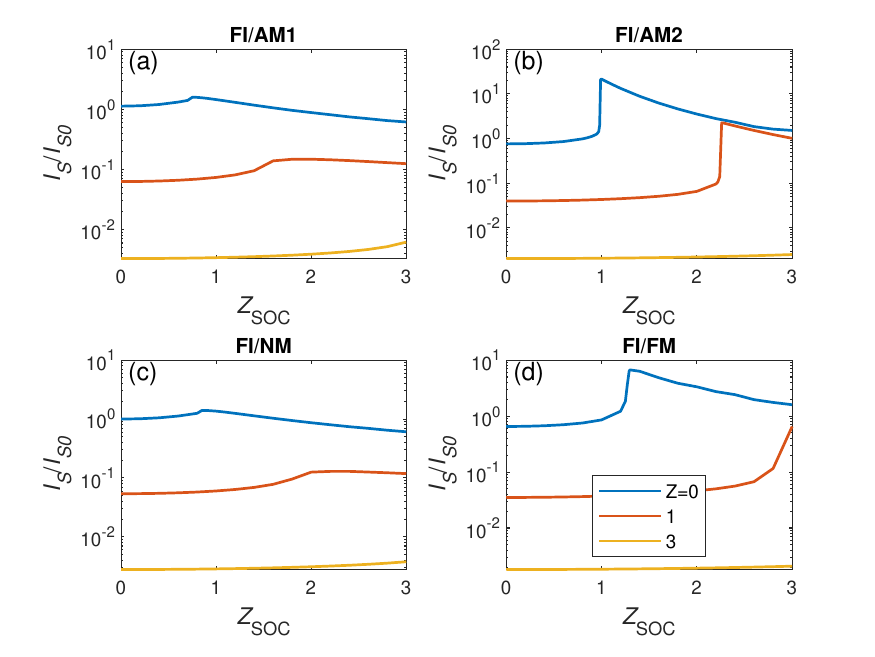}
	\caption{(Color online) Normalized spin pumping current $I_s/I_{s0}$ as a function of Rashba $Z_\text{SOC}$ for FI/AM1 and FI/AM2 in (a) and (b), respectively, in which $\alpha/\alpha_c=0.6$. (c) $I_s/I_{s0}$ as a function of $Z_\text{SOC}$ for FI/NM. (d)  $I_s/I_{s0}$ as a function of $Z_\text{SOC}$ for FI/FM with $J_\text{ex}/\mu_0=0.6$. Different interfacial barriers $Z=0,1,3$ are considered. Here $m=0.2$ and $\hbar\Omega=0.5$ meV are utilized. $I_{s0}$ corresponds to the spin pumping current for FI/NM with $\mu/\mu_0=1$ in the absence of Rashba SOC, the same as $I_{s0}$ used in Fig. \ref{fig:alpha}.
	}
	\label{fig:zsoc}
\end{figure}

\textit{Results: Spin-orbit dependence. --}  Similar to the barrier $Z= \frac{m_e U_0}{\hbar^2 k_F}$, the interfacial Rashba SOC can be characterized by introducing the dimensionless parameter $Z_\text{SOC}=\frac{m_eU_\text{SO}}{\hbar^2k_F}$, based on which $k_{\alpha,\sigma}$ in Eq. (\ref{eq:BC_main}) can be written as $k_{\alpha,\sigma}=2Zk_F-2Z_\text{SOC}k_y\sigma-\i\frac{\alpha m_e k_y}{\hbar^2}\sigma$ with $\sigma=+1(-1)$. In Fig. \ref{fig:zsoc}, the spin pumping current is plotted as a function of $Z_\text{SOC}$ for different bilayers with gradually increasing interface barrier $Z=0,1,3$. A non-monotonic behavior with a maximum whose position can be shifted with $Z$ is achieved in all setups. This is related to the effective spin-dependent barrier induced by SOC in the form of $k_y\sigma$ in $k_{\alpha,\sigma}$. When $Z_\text{SOC}$ is present and $Z$ is fixed, there exists an optimal value of $Z_\text{SOC}$ where the barrier is strongly reduced for many angles of incidence (i.e., $k_y$ modes) of a given spin type due to the $k_y\sigma$ dependence in the boundary condition, resulting in enhanced spin-flip and spin current. When $Z_\text{SOC}$ continues to increase, the total barrier then increases again which causes less spin-flip and reduces the spin current. Note that the Fermi-level mismatch between the two layers also results in normal reflection and acts as an effective barrier even when $Z=0$ \cite{mismatch_barrier_1}, which can thus be compensated by $Z_\text{SOC}$ to achieve the optimal spin current via the argument above. 

In the absence of $Z_\text{SOC}$, it is shown in Fig. \ref{fig:alpha} that FI/AM1 produces a larger spin pumping current compared with FI/AM2, indicating that AM1 is the spin pumping-enhanced-orientation. However, this changes when $Z_\text{SOC}$ is present. FI/AM2 with the spin pumping-suppressed-orientation can in that case generate a much larger spin current compared with FI/AM1 when $Z_\text{SOC}$ is tuned to its optimal value, as shown in Fig. \ref{fig:zsoc}(b). Similar behavior can be observed in FI/FM [Fig. \ref{fig:zsoc}(d)] but with a smaller spin pumping current maximum compared with FI/AM2. The suppression of spin current due to interfacial Rashba interaction via spin memory loss and spin current absorption has been studied previously \cite{SP_SOC_Shufeng} within a perturbative framework.

\textit{Concluding remarks. --} We investigate spin pumping from a FI to an AM by considering two representative AMs with 0 and 45-degree rotation relative to the interface. We find the spin pumping current can be both enhanced and suppressed by altermagnetism depending on the interface orientation. In addition, the inclusion of interfacial Rashba SOC strongly affects the spin pumping current by changing the preferred interface orientation for altermagnetism when the SOC strength is optimized, indicating the crucial role of the interfacial properties for spin pumping in altermagnets.

\textit{Acknowledgments. --} This work was supported by the Research
Council of Norway through Grant No. 323766 and its Centres
of Excellence funding scheme Grant No. 262633 “QuSpin”. Support from
Sigma2 - the National Infrastructure for High Performance
Computing and Data Storage in Norway, project NN9577K, is acknowledged.

\bibliography{main.bib}

\begin{thebibliography}{43}%
\makeatletter
\providecommand \@ifxundefined [1]{%
 \@ifx{#1\undefined}
}%
\providecommand \@ifnum [1]{%
 \ifnum #1\expandafter \@firstoftwo
 \else \expandafter \@secondoftwo
 \fi
}%
\providecommand \@ifx [1]{%
 \ifx #1\expandafter \@firstoftwo
 \else \expandafter \@secondoftwo
 \fi
}%
\providecommand \natexlab [1]{#1}%
\providecommand \enquote  [1]{``#1''}%
\providecommand \bibnamefont  [1]{#1}%
\providecommand \bibfnamefont [1]{#1}%
\providecommand \citenamefont [1]{#1}%
\providecommand \href@noop [0]{\@secondoftwo}%
\providecommand \href [0]{\begingroup \@sanitize@url \@href}%
\providecommand \@href[1]{\@@startlink{#1}\@@href}%
\providecommand \@@href[1]{\endgroup#1\@@endlink}%
\providecommand \@sanitize@url [0]{\catcode `\\12\catcode `\$12\catcode
  `\&12\catcode `\#12\catcode `\^12\catcode `\_12\catcode `\%12\relax}%
\providecommand \@@startlink[1]{}%
\providecommand \@@endlink[0]{}%
\providecommand \url  [0]{\begingroup\@sanitize@url \@url }%
\providecommand \@url [1]{\endgroup\@href {#1}{\urlprefix }}%
\providecommand \urlprefix  [0]{URL }%
\providecommand \Eprint [0]{\href }%
\providecommand \doibase [0]{https://doi.org/}%
\providecommand \selectlanguage [0]{\@gobble}%
\providecommand \bibinfo  [0]{\@secondoftwo}%
\providecommand \bibfield  [0]{\@secondoftwo}%
\providecommand \translation [1]{[#1]}%
\providecommand \BibitemOpen [0]{}%
\providecommand \bibitemStop [0]{}%
\providecommand \bibitemNoStop [0]{.\EOS\space}%
\providecommand \EOS [0]{\spacefactor3000\relax}%
\providecommand \BibitemShut  [1]{\csname bibitem#1\endcsname}%
\let\auto@bib@innerbib\@empty
%</preamble>
\bibitem [{\citenamefont {Tserkovnyak}\ \emph {et~al.}(2005)\citenamefont
  {Tserkovnyak}, \citenamefont {Brataas}, \citenamefont {Bauer},\ and\
  \citenamefont {Halperin}}]{tserkovnyak_rmp_04}%
  \BibitemOpen
  \bibfield  {author} {\bibinfo {author} {\bibfnamefont {Y.}~\bibnamefont
  {Tserkovnyak}}, \bibinfo {author} {\bibfnamefont {A.}~\bibnamefont
  {Brataas}}, \bibinfo {author} {\bibfnamefont {G.~E.~W.}\ \bibnamefont
  {Bauer}},\ and\ \bibinfo {author} {\bibfnamefont {B.~I.}\ \bibnamefont
  {Halperin}},\ }\href {https://doi.org/10.1103/RevModPhys.77.1375} {\bibfield
  {journal} {\bibinfo  {journal} {Rev. Mod. Phys.}\ }\textbf {\bibinfo {volume}
  {77}},\ \bibinfo {pages} {1375} (\bibinfo {year} {2005})}\BibitemShut
  {NoStop}%
\bibitem [{\citenamefont {Tserkovnyak}\ \emph
  {et~al.}(2002{\natexlab{a}})\citenamefont {Tserkovnyak}, \citenamefont
  {Brataas},\ and\ \citenamefont {Bauer}}]{Tserkovnyak2002Dec}%
  \BibitemOpen
  \bibfield  {author} {\bibinfo {author} {\bibfnamefont {Y.}~\bibnamefont
  {Tserkovnyak}}, \bibinfo {author} {\bibfnamefont {A.}~\bibnamefont
  {Brataas}},\ and\ \bibinfo {author} {\bibfnamefont {G.~E.~W.}\ \bibnamefont
  {Bauer}},\ }\href {https://doi.org/10.1103/PhysRevB.66.224403} {\bibfield
  {journal} {\bibinfo  {journal} {Phys. Rev. B}\ }\textbf {\bibinfo {volume}
  {66}},\ \bibinfo {pages} {224403} (\bibinfo {year}
  {2002}{\natexlab{a}})}\BibitemShut {NoStop}%
\bibitem [{\citenamefont {Tserkovnyak}\ \emph
  {et~al.}(2002{\natexlab{b}})\citenamefont {Tserkovnyak}, \citenamefont
  {Brataas},\ and\ \citenamefont {Bauer}}]{Tserkovnyak2002Feb}%
  \BibitemOpen
  \bibfield  {author} {\bibinfo {author} {\bibfnamefont {Y.}~\bibnamefont
  {Tserkovnyak}}, \bibinfo {author} {\bibfnamefont {A.}~\bibnamefont
  {Brataas}},\ and\ \bibinfo {author} {\bibfnamefont {G.~E.~W.}\ \bibnamefont
  {Bauer}},\ }\href {https://doi.org/10.1103/PhysRevLett.88.117601} {\bibfield
  {journal} {\bibinfo  {journal} {Phys. Rev. Lett.}\ }\textbf {\bibinfo
  {volume} {88}},\ \bibinfo {pages} {117601} (\bibinfo {year}
  {2002}{\natexlab{b}})}\BibitemShut {NoStop}%
\bibitem [{\citenamefont {Brataas}\ \emph {et~al.}(2020)\citenamefont
  {Brataas}, \citenamefont {van Wees}, \citenamefont {Klein}, \citenamefont
  {de~Loubens},\ and\ \citenamefont {Viret}}]{Brataas2020Nov}%
  \BibitemOpen
  \bibfield  {author} {\bibinfo {author} {\bibfnamefont {A.}~\bibnamefont
  {Brataas}}, \bibinfo {author} {\bibfnamefont {B.}~\bibnamefont {van Wees}},
  \bibinfo {author} {\bibfnamefont {O.}~\bibnamefont {Klein}}, \bibinfo
  {author} {\bibfnamefont {G.}~\bibnamefont {de~Loubens}},\ and\ \bibinfo
  {author} {\bibfnamefont {M.}~\bibnamefont {Viret}},\ }\href
  {https://doi.org/10.1016/j.physrep.2020.08.006} {\bibfield  {journal}
  {\bibinfo  {journal} {Phys. Rep.}\ }\textbf {\bibinfo {volume} {885}},\
  \bibinfo {pages} {1} (\bibinfo {year} {2020})}\BibitemShut {NoStop}%
\bibitem [{\citenamefont {Sun}\ \emph {et~al.}(2022)\citenamefont {Sun},
  \citenamefont {Yang},\ and\ \citenamefont {Jalil}}]{Sun2022Mar}%
  \BibitemOpen
  \bibfield  {author} {\bibinfo {author} {\bibfnamefont {C.}~\bibnamefont
  {Sun}}, \bibinfo {author} {\bibfnamefont {H.}~\bibnamefont {Yang}},\ and\
  \bibinfo {author} {\bibfnamefont {M.~B.~A.}\ \bibnamefont {Jalil}},\ }\href
  {https://doi.org/10.1103/PhysRevB.105.104407} {\bibfield  {journal} {\bibinfo
   {journal} {Phys. Rev. B}\ }\textbf {\bibinfo {volume} {105}},\ \bibinfo
  {pages} {104407} (\bibinfo {year} {2022})}\BibitemShut {NoStop}%
\bibitem [{\citenamefont {Heinrich}\ \emph {et~al.}(2011)\citenamefont
  {Heinrich}, \citenamefont {Burrowes}, \citenamefont {Montoya}, \citenamefont
  {Kardasz}, \citenamefont {Girt}, \citenamefont {Song}, \citenamefont {Sun},\
  and\ \citenamefont {Wu}}]{Heinrich2011Aug}%
  \BibitemOpen
  \bibfield  {author} {\bibinfo {author} {\bibfnamefont {B.}~\bibnamefont
  {Heinrich}}, \bibinfo {author} {\bibfnamefont {C.}~\bibnamefont {Burrowes}},
  \bibinfo {author} {\bibfnamefont {E.}~\bibnamefont {Montoya}}, \bibinfo
  {author} {\bibfnamefont {B.}~\bibnamefont {Kardasz}}, \bibinfo {author}
  {\bibfnamefont {E.}~\bibnamefont {Girt}}, \bibinfo {author} {\bibfnamefont
  {Y.-Y.}\ \bibnamefont {Song}}, \bibinfo {author} {\bibfnamefont
  {Y.}~\bibnamefont {Sun}},\ and\ \bibinfo {author} {\bibfnamefont
  {M.}~\bibnamefont {Wu}},\ }\href
  {https://doi.org/10.1103/PhysRevLett.107.066604} {\bibfield  {journal}
  {\bibinfo  {journal} {Phys. Rev. Lett.}\ }\textbf {\bibinfo {volume} {107}},\
  \bibinfo {pages} {066604} (\bibinfo {year} {2011})}\BibitemShut {NoStop}%
\bibitem [{\citenamefont {Haertinger}\ \emph {et~al.}(2015)\citenamefont
  {Haertinger}, \citenamefont {Back}, \citenamefont {Lotze}, \citenamefont
  {Weiler}, \citenamefont {Gepr{\ifmmode\ddot{a}\else\"{a}\fi}gs},
  \citenamefont {Huebl}, \citenamefont {Goennenwein},\ and\ \citenamefont
  {Woltersdorf}}]{Haertinger2015Aug}%
  \BibitemOpen
  \bibfield  {author} {\bibinfo {author} {\bibfnamefont {M.}~\bibnamefont
  {Haertinger}}, \bibinfo {author} {\bibfnamefont {C.~H.}\ \bibnamefont
  {Back}}, \bibinfo {author} {\bibfnamefont {J.}~\bibnamefont {Lotze}},
  \bibinfo {author} {\bibfnamefont {M.}~\bibnamefont {Weiler}}, \bibinfo
  {author} {\bibfnamefont {S.}~\bibnamefont
  {Gepr{\ifmmode\ddot{a}\else\"{a}\fi}gs}}, \bibinfo {author} {\bibfnamefont
  {H.}~\bibnamefont {Huebl}}, \bibinfo {author} {\bibfnamefont {S.~T.~B.}\
  \bibnamefont {Goennenwein}},\ and\ \bibinfo {author} {\bibfnamefont
  {G.}~\bibnamefont {Woltersdorf}},\ }\href
  {https://doi.org/10.1103/PhysRevB.92.054437} {\bibfield  {journal} {\bibinfo
  {journal} {Phys. Rev. B}\ }\textbf {\bibinfo {volume} {92}},\ \bibinfo
  {pages} {054437} (\bibinfo {year} {2015})}\BibitemShut {NoStop}%
\bibitem [{\citenamefont {Yang}\ and\ \citenamefont {Hammel}(2018)}]{fmr_sp}%
  \BibitemOpen
  \bibfield  {author} {\bibinfo {author} {\bibfnamefont {F.}~\bibnamefont
  {Yang}}\ and\ \bibinfo {author} {\bibfnamefont {P.~C.}\ \bibnamefont
  {Hammel}},\ }\href {https://doi.org/10.1088/1361-6463/aac249} {\bibfield
  {journal} {\bibinfo  {journal} {J. Phys. D: Appl. Phys.}\ }\textbf {\bibinfo
  {volume} {51}},\ \bibinfo {pages} {253001} (\bibinfo {year}
  {2018})}\BibitemShut {NoStop}%
\bibitem [{\citenamefont {Rezende}\ \emph {et~al.}(2013)\citenamefont
  {Rezende}, \citenamefont {Rodriguez-Suarez}, \citenamefont {Soares},
  \citenamefont {Vilela-Leao}, \citenamefont {Ley~Dominguez},\ and\
  \citenamefont {Azevedo}}]{Rezende2013enhanced}%
  \BibitemOpen
  \bibfield  {author} {\bibinfo {author} {\bibfnamefont {S.~M.}\ \bibnamefont
  {Rezende}}, \bibinfo {author} {\bibfnamefont {R.~L.}\ \bibnamefont
  {Rodriguez-Suarez}}, \bibinfo {author} {\bibfnamefont {M.~M.}\ \bibnamefont
  {Soares}}, \bibinfo {author} {\bibfnamefont {L.~H.}\ \bibnamefont
  {Vilela-Leao}}, \bibinfo {author} {\bibfnamefont {D.}~\bibnamefont
  {Ley~Dominguez}},\ and\ \bibinfo {author} {\bibfnamefont {A.}~\bibnamefont
  {Azevedo}},\ }\bibfield  {journal} {\bibinfo  {journal} {Appl. Phys. Lett.}\
  }\textbf {\bibinfo {volume} {102}},\ \href
  {https://doi.org/10.1063/1.4773993} {10.1063/1.4773993} (\bibinfo {year}
  {2013})\BibitemShut {NoStop}%
\bibitem [{\citenamefont {Castel}\ \emph {et~al.}(2012)\citenamefont {Castel},
  \citenamefont {Vlietstra}, \citenamefont {van Wees},\ and\ \citenamefont
  {Youssef}}]{Castel2012Oct}%
  \BibitemOpen
  \bibfield  {author} {\bibinfo {author} {\bibfnamefont {V.}~\bibnamefont
  {Castel}}, \bibinfo {author} {\bibfnamefont {N.}~\bibnamefont {Vlietstra}},
  \bibinfo {author} {\bibfnamefont {B.~J.}\ \bibnamefont {van Wees}},\ and\
  \bibinfo {author} {\bibfnamefont {J.~B.}\ \bibnamefont {Youssef}},\ }\href
  {https://doi.org/10.1103/PhysRevB.86.134419} {\bibfield  {journal} {\bibinfo
  {journal} {Phys. Rev. B}\ }\textbf {\bibinfo {volume} {86}},\ \bibinfo
  {pages} {134419} (\bibinfo {year} {2012})}\BibitemShut {NoStop}%
\bibitem [{\citenamefont {Sun}\ and\ \citenamefont
  {Linder}(2023)}]{Sun2023Apr}%
  \BibitemOpen
  \bibfield  {author} {\bibinfo {author} {\bibfnamefont {C.}~\bibnamefont
  {Sun}}\ and\ \bibinfo {author} {\bibfnamefont {J.}~\bibnamefont {Linder}},\
  }\href {https://doi.org/10.1103/PhysRevB.107.144504} {\bibfield  {journal}
  {\bibinfo  {journal} {Phys. Rev. B}\ }\textbf {\bibinfo {volume} {107}},\
  \bibinfo {pages} {144504} (\bibinfo {year} {2023})}\BibitemShut {NoStop}%
\bibitem [{\citenamefont {Jeon}\ \emph {et~al.}(2018)\citenamefont {Jeon},
  \citenamefont {Ciccarelli}, \citenamefont {Ferguson}, \citenamefont
  {Kurebayashi}, \citenamefont {Cohen}, \citenamefont {Montiel}, \citenamefont
  {Eschrig}, \citenamefont {Robinson},\ and\ \citenamefont
  {Blamire}}]{Jeon2018Jun}%
  \BibitemOpen
  \bibfield  {author} {\bibinfo {author} {\bibfnamefont {K.-R.}\ \bibnamefont
  {Jeon}}, \bibinfo {author} {\bibfnamefont {C.}~\bibnamefont {Ciccarelli}},
  \bibinfo {author} {\bibfnamefont {A.~J.}\ \bibnamefont {Ferguson}}, \bibinfo
  {author} {\bibfnamefont {H.}~\bibnamefont {Kurebayashi}}, \bibinfo {author}
  {\bibfnamefont {L.~F.}\ \bibnamefont {Cohen}}, \bibinfo {author}
  {\bibfnamefont {X.}~\bibnamefont {Montiel}}, \bibinfo {author} {\bibfnamefont
  {M.}~\bibnamefont {Eschrig}}, \bibinfo {author} {\bibfnamefont {J.~W.~A.}\
  \bibnamefont {Robinson}},\ and\ \bibinfo {author} {\bibfnamefont {M.~G.}\
  \bibnamefont {Blamire}},\ }\href {https://doi.org/10.1038/s41563-018-0058-9}
  {\bibfield  {journal} {\bibinfo  {journal} {Nat. Mater.}\ }\textbf {\bibinfo
  {volume} {17}},\ \bibinfo {pages} {499} (\bibinfo {year} {2018})}\BibitemShut
  {NoStop}%
\bibitem [{\citenamefont {Yao}\ \emph {et~al.}(2018)\citenamefont {Yao},
  \citenamefont {Song}, \citenamefont {Takamura}, \citenamefont {Cascales},
  \citenamefont {Yuan}, \citenamefont {Ma}, \citenamefont {Yun}, \citenamefont
  {Xie}, \citenamefont {Moodera},\ and\ \citenamefont {Han}}]{Yao2018Jun}%
  \BibitemOpen
  \bibfield  {author} {\bibinfo {author} {\bibfnamefont {Y.}~\bibnamefont
  {Yao}}, \bibinfo {author} {\bibfnamefont {Q.}~\bibnamefont {Song}}, \bibinfo
  {author} {\bibfnamefont {Y.}~\bibnamefont {Takamura}}, \bibinfo {author}
  {\bibfnamefont {J.~P.}\ \bibnamefont {Cascales}}, \bibinfo {author}
  {\bibfnamefont {W.}~\bibnamefont {Yuan}}, \bibinfo {author} {\bibfnamefont
  {Y.}~\bibnamefont {Ma}}, \bibinfo {author} {\bibfnamefont {Y.}~\bibnamefont
  {Yun}}, \bibinfo {author} {\bibfnamefont {X.~C.}\ \bibnamefont {Xie}},
  \bibinfo {author} {\bibfnamefont {J.~S.}\ \bibnamefont {Moodera}},\ and\
  \bibinfo {author} {\bibfnamefont {W.}~\bibnamefont {Han}},\ }\href
  {https://doi.org/10.1103/PhysRevB.97.224414} {\bibfield  {journal} {\bibinfo
  {journal} {Phys. Rev. B}\ }\textbf {\bibinfo {volume} {97}},\ \bibinfo
  {pages} {224414} (\bibinfo {year} {2018})}\BibitemShut {NoStop}%
\bibitem [{\citenamefont {Carreira}\ \emph {et~al.}(2021)\citenamefont
  {Carreira}, \citenamefont {Sanchez-Manzano}, \citenamefont {Yoo},
  \citenamefont {Seurre}, \citenamefont {Rouco}, \citenamefont {Sander},
  \citenamefont {Santamaria}, \citenamefont {Anane},\ and\ \citenamefont
  {Villegas}}]{Carreira2021Oct}%
  \BibitemOpen
  \bibfield  {author} {\bibinfo {author} {\bibfnamefont {S.~J.}\ \bibnamefont
  {Carreira}}, \bibinfo {author} {\bibfnamefont {D.}~\bibnamefont
  {Sanchez-Manzano}}, \bibinfo {author} {\bibfnamefont {M.-W.}\ \bibnamefont
  {Yoo}}, \bibinfo {author} {\bibfnamefont {K.}~\bibnamefont {Seurre}},
  \bibinfo {author} {\bibfnamefont {V.}~\bibnamefont {Rouco}}, \bibinfo
  {author} {\bibfnamefont {A.}~\bibnamefont {Sander}}, \bibinfo {author}
  {\bibfnamefont {J.}~\bibnamefont {Santamaria}}, \bibinfo {author}
  {\bibfnamefont {A.}~\bibnamefont {Anane}},\ and\ \bibinfo {author}
  {\bibfnamefont {J.~E.}\ \bibnamefont {Villegas}},\ }\href
  {https://doi.org/10.1103/PhysRevB.104.144428} {\bibfield  {journal} {\bibinfo
   {journal} {Phys. Rev. B}\ }\textbf {\bibinfo {volume} {104}},\ \bibinfo
  {pages} {144428} (\bibinfo {year} {2021})}\BibitemShut {NoStop}%
\bibitem [{\citenamefont {Inoue}\ \emph {et~al.}(2017)\citenamefont {Inoue},
  \citenamefont {Ichioka},\ and\ \citenamefont {Adachi}}]{inoue_17}%
  \BibitemOpen
  \bibfield  {author} {\bibinfo {author} {\bibfnamefont {M.}~\bibnamefont
  {Inoue}}, \bibinfo {author} {\bibfnamefont {M.}~\bibnamefont {Ichioka}},\
  and\ \bibinfo {author} {\bibfnamefont {H.}~\bibnamefont {Adachi}},\ }\href
  {https://doi.org/10.1103/PhysRevB.96.024414} {\bibfield  {journal} {\bibinfo
  {journal} {Phys. Rev. B}\ }\textbf {\bibinfo {volume} {96}},\ \bibinfo
  {pages} {024414} (\bibinfo {year} {2017})}\BibitemShut {NoStop}%
\bibitem [{\citenamefont {Kato}\ \emph {et~al.}(2019)\citenamefont {Kato},
  \citenamefont {Ohnuma}, \citenamefont {Matsuo}, \citenamefont {Rech},
  \citenamefont {Jonckheere},\ and\ \citenamefont {Martin}}]{kato_prb_2019}%
  \BibitemOpen
  \bibfield  {author} {\bibinfo {author} {\bibfnamefont {T.}~\bibnamefont
  {Kato}}, \bibinfo {author} {\bibfnamefont {Y.}~\bibnamefont {Ohnuma}},
  \bibinfo {author} {\bibfnamefont {M.}~\bibnamefont {Matsuo}}, \bibinfo
  {author} {\bibfnamefont {J.}~\bibnamefont {Rech}}, \bibinfo {author}
  {\bibfnamefont {T.}~\bibnamefont {Jonckheere}},\ and\ \bibinfo {author}
  {\bibfnamefont {T.}~\bibnamefont {Martin}},\ }\href
  {https://doi.org/10.1103/PhysRevB.99.144411} {\bibfield  {journal} {\bibinfo
  {journal} {Phys. Rev. B}\ }\textbf {\bibinfo {volume} {99}},\ \bibinfo
  {pages} {144411} (\bibinfo {year} {2019})}\BibitemShut {NoStop}%
\bibitem [{\citenamefont {Ominato}\ \emph
  {et~al.}(2022{\natexlab{a}})\citenamefont {Ominato}, \citenamefont
  {Yamakage}, \citenamefont {Kato},\ and\ \citenamefont
  {Matsuo}}]{ominato_prb_2022}%
  \BibitemOpen
  \bibfield  {author} {\bibinfo {author} {\bibfnamefont {Y.}~\bibnamefont
  {Ominato}}, \bibinfo {author} {\bibfnamefont {A.}~\bibnamefont {Yamakage}},
  \bibinfo {author} {\bibfnamefont {T.}~\bibnamefont {Kato}},\ and\ \bibinfo
  {author} {\bibfnamefont {M.}~\bibnamefont {Matsuo}},\ }\href
  {https://doi.org/10.1103/PhysRevB.105.205406} {\bibfield  {journal} {\bibinfo
   {journal} {Phys. Rev. B}\ }\textbf {\bibinfo {volume} {105}},\ \bibinfo
  {pages} {205406} (\bibinfo {year} {2022}{\natexlab{a}})}\BibitemShut
  {NoStop}%
\bibitem [{\citenamefont {Ominato}\ \emph
  {et~al.}(2022{\natexlab{b}})\citenamefont {Ominato}, \citenamefont
  {Yamakage},\ and\ \citenamefont {Matsuo}}]{ominato_prb_2022a}%
  \BibitemOpen
  \bibfield  {author} {\bibinfo {author} {\bibfnamefont {Y.}~\bibnamefont
  {Ominato}}, \bibinfo {author} {\bibfnamefont {A.}~\bibnamefont {Yamakage}},\
  and\ \bibinfo {author} {\bibfnamefont {M.}~\bibnamefont {Matsuo}},\ }\href
  {https://doi.org/10.1103/PhysRevB.106.L161406} {\bibfield  {journal}
  {\bibinfo  {journal} {Phys. Rev. B}\ }\textbf {\bibinfo {volume} {106}},\
  \bibinfo {pages} {L161406} (\bibinfo {year}
  {2022}{\natexlab{b}})}\BibitemShut {NoStop}%
\bibitem [{\citenamefont {Ahn}\ \emph {et~al.}(2019)\citenamefont {Ahn},
  \citenamefont {Hariki}, \citenamefont {Lee},\ and\ \citenamefont
  {Kune{\ifmmode\check{s}\else\v{s}\fi}}}]{ahn_prb_2019}%
  \BibitemOpen
  \bibfield  {author} {\bibinfo {author} {\bibfnamefont {K.-H.}\ \bibnamefont
  {Ahn}}, \bibinfo {author} {\bibfnamefont {A.}~\bibnamefont {Hariki}},
  \bibinfo {author} {\bibfnamefont {K.-W.}\ \bibnamefont {Lee}},\ and\ \bibinfo
  {author} {\bibfnamefont {J.}~\bibnamefont
  {Kune{\ifmmode\check{s}\else\v{s}\fi}}},\ }\href
  {https://doi.org/10.1103/PhysRevB.99.184432} {\bibfield  {journal} {\bibinfo
  {journal} {Phys. Rev. B}\ }\textbf {\bibinfo {volume} {99}},\ \bibinfo
  {pages} {184432} (\bibinfo {year} {2019})}\BibitemShut {NoStop}%
\bibitem [{\citenamefont {Hayami}\ \emph {et~al.}(2019)\citenamefont {Hayami},
  \citenamefont {Yanagi},\ and\ \citenamefont {Kusunose}}]{hayami_jpsj_19}%
  \BibitemOpen
  \bibfield  {author} {\bibinfo {author} {\bibfnamefont {S.}~\bibnamefont
  {Hayami}}, \bibinfo {author} {\bibfnamefont {Y.}~\bibnamefont {Yanagi}},\
  and\ \bibinfo {author} {\bibfnamefont {H.}~\bibnamefont {Kusunose}},\
  }\href@noop {} {\bibfield  {journal} {\bibinfo  {journal} {J. Phys. Soc.
  Jpn.}\ }\textbf {\bibinfo {volume} {88}},\ \bibinfo {pages} {123702}
  (\bibinfo {year} {2019})}\BibitemShut {NoStop}%
\bibitem [{\citenamefont {{\v S}mejkal}\ \emph {et~al.}(2020)\citenamefont {{\v
  S}mejkal}, \citenamefont {{Gonz{\'a}lez-Hern{\'a}ndez}}, \citenamefont
  {Jungwirth},\ and\ \citenamefont {Sinova}}]{smejkal_sa_20}%
  \BibitemOpen
  \bibfield  {author} {\bibinfo {author} {\bibfnamefont {L.}~\bibnamefont {{\v
  S}mejkal}}, \bibinfo {author} {\bibfnamefont {R.}~\bibnamefont
  {{Gonz{\'a}lez-Hern{\'a}ndez}}}, \bibinfo {author} {\bibfnamefont
  {T.}~\bibnamefont {Jungwirth}},\ and\ \bibinfo {author} {\bibfnamefont
  {J.}~\bibnamefont {Sinova}},\ }\href {https://doi.org/10.1126/sciadv.aaz8809}
  {\bibfield  {journal} {\bibinfo  {journal} {Sci. Adv.}\ }\textbf {\bibinfo
  {volume} {6}},\ \bibinfo {pages} {eaaz8809} (\bibinfo {year}
  {2020})}\BibitemShut {NoStop}%
\bibitem [{\citenamefont {Yuan}\ \emph {et~al.}(2020)\citenamefont {Yuan},
  \citenamefont {Wang}, \citenamefont {Luo}, \citenamefont {Rashba},\ and\
  \citenamefont {Zunger}}]{yuan_prb_20}%
  \BibitemOpen
  \bibfield  {author} {\bibinfo {author} {\bibfnamefont {L.-D.}\ \bibnamefont
  {Yuan}}, \bibinfo {author} {\bibfnamefont {Z.}~\bibnamefont {Wang}}, \bibinfo
  {author} {\bibfnamefont {J.-W.}\ \bibnamefont {Luo}}, \bibinfo {author}
  {\bibfnamefont {E.~I.}\ \bibnamefont {Rashba}},\ and\ \bibinfo {author}
  {\bibfnamefont {A.}~\bibnamefont {Zunger}},\ }\href
  {https://doi.org/10.1103/PhysRevB.102.014422} {\bibfield  {journal} {\bibinfo
   {journal} {Phys. Rev. B}\ }\textbf {\bibinfo {volume} {102}},\ \bibinfo
  {pages} {014422} (\bibinfo {year} {2020})}\BibitemShut {NoStop}%
\bibitem [{\citenamefont {{\ifmmode\check{S}\else\v{S}\fi}mejkal}\ \emph
  {et~al.}(2022{\natexlab{a}})\citenamefont
  {{\ifmmode\check{S}\else\v{S}\fi}mejkal}, \citenamefont {Sinova},\ and\
  \citenamefont {Jungwirth}}]{am_emerging_22}%
  \BibitemOpen
  \bibfield  {author} {\bibinfo {author} {\bibfnamefont {L.}~\bibnamefont
  {{\ifmmode\check{S}\else\v{S}\fi}mejkal}}, \bibinfo {author} {\bibfnamefont
  {J.}~\bibnamefont {Sinova}},\ and\ \bibinfo {author} {\bibfnamefont
  {T.}~\bibnamefont {Jungwirth}},\ }\href
  {https://doi.org/10.1103/PhysRevX.12.040501} {\bibfield  {journal} {\bibinfo
  {journal} {Phys. Rev. X}\ }\textbf {\bibinfo {volume} {12}},\ \bibinfo
  {pages} {040501} (\bibinfo {year} {2022}{\natexlab{a}})}\BibitemShut
  {NoStop}%
\bibitem [{\citenamefont {{\ifmmode\check{S}\else\v{S}\fi}mejkal}\ \emph
  {et~al.}(2022{\natexlab{b}})\citenamefont
  {{\ifmmode\check{S}\else\v{S}\fi}mejkal}, \citenamefont {Sinova},\ and\
  \citenamefont {Jungwirth}}]{vsmejkal_2022beyond}%
  \BibitemOpen
  \bibfield  {author} {\bibinfo {author} {\bibfnamefont {L.}~\bibnamefont
  {{\ifmmode\check{S}\else\v{S}\fi}mejkal}}, \bibinfo {author} {\bibfnamefont
  {J.}~\bibnamefont {Sinova}},\ and\ \bibinfo {author} {\bibfnamefont
  {T.}~\bibnamefont {Jungwirth}},\ }\href
  {https://doi.org/10.1103/PhysRevX.12.031042} {\bibfield  {journal} {\bibinfo
  {journal} {Phys. Rev. X}\ }\textbf {\bibinfo {volume} {12}},\ \bibinfo
  {pages} {031042} (\bibinfo {year} {2022}{\natexlab{b}})}\BibitemShut
  {NoStop}%
\bibitem [{\citenamefont {Mazin}\ and\ \citenamefont
  {Editors}(2022)}]{mazin2022_altermagnetism}%
  \BibitemOpen
  \bibfield  {author} {\bibinfo {author} {\bibfnamefont {I.}~\bibnamefont
  {Mazin}}\ and\ \bibinfo {author} {\bibfnamefont {T.~P.}\ \bibnamefont
  {Editors}},\ }\href {https://doi.org/10.1103/PhysRevX.12.040002} {\bibfield
  {journal} {\bibinfo  {journal} {Phys. Rev. X}\ }\textbf {\bibinfo {volume}
  {12}},\ \bibinfo {pages} {040002} (\bibinfo {year} {2022})}\BibitemShut
  {NoStop}%
\bibitem [{\citenamefont {Mazin}(2023)}]{mazin2023_altermagnetism}%
  \BibitemOpen
  \bibfield  {author} {\bibinfo {author} {\bibfnamefont {I.~I.}\ \bibnamefont
  {Mazin}},\ }\href {https://doi.org/10.1103/PhysRevB.107.L100418} {\bibfield
  {journal} {\bibinfo  {journal} {Phys. Rev. B}\ }\textbf {\bibinfo {volume}
  {107}},\ \bibinfo {pages} {L100418} (\bibinfo {year} {2023})}\BibitemShut
  {NoStop}%
\bibitem [{\citenamefont {Chen}\ and\ \citenamefont
  {Zhang}(2015)}]{SP_SOC_Shufeng}%
  \BibitemOpen
  \bibfield  {author} {\bibinfo {author} {\bibfnamefont {K.}~\bibnamefont
  {Chen}}\ and\ \bibinfo {author} {\bibfnamefont {S.}~\bibnamefont {Zhang}},\
  }\href {https://doi.org/10.1103/PhysRevLett.114.126602} {\bibfield  {journal}
  {\bibinfo  {journal} {Phys. Rev. Lett.}\ }\textbf {\bibinfo {volume} {114}},\
  \bibinfo {pages} {126602} (\bibinfo {year} {2015})}\BibitemShut {NoStop}%
\bibitem [{\citenamefont {{\ifmmode\check{S}\else\v{S}\fi}mejkal}\ \emph
  {et~al.}(2022{\natexlab{c}})\citenamefont
  {{\ifmmode\check{S}\else\v{S}\fi}mejkal}, \citenamefont {Marmodoro},
  \citenamefont {Ahn}, \citenamefont {Gonzalez-Hernandez}, \citenamefont
  {Turek}, \citenamefont {Mankovsky}, \citenamefont {Ebert}, \citenamefont
  {D'Souza}, \citenamefont {{\ifmmode\check{S}\else\v{S}\fi}ipr}, \citenamefont
  {Sinova},\ and\ \citenamefont {Jungwirth}}]{RuO2_22}%
  \BibitemOpen
  \bibfield  {author} {\bibinfo {author} {\bibfnamefont {L.}~\bibnamefont
  {{\ifmmode\check{S}\else\v{S}\fi}mejkal}}, \bibinfo {author} {\bibfnamefont
  {A.}~\bibnamefont {Marmodoro}}, \bibinfo {author} {\bibfnamefont {K.-H.}\
  \bibnamefont {Ahn}}, \bibinfo {author} {\bibfnamefont {R.}~\bibnamefont
  {Gonzalez-Hernandez}}, \bibinfo {author} {\bibfnamefont {I.}~\bibnamefont
  {Turek}}, \bibinfo {author} {\bibfnamefont {S.}~\bibnamefont {Mankovsky}},
  \bibinfo {author} {\bibfnamefont {H.}~\bibnamefont {Ebert}}, \bibinfo
  {author} {\bibfnamefont {S.~W.}\ \bibnamefont {D'Souza}}, \bibinfo {author}
  {\bibfnamefont {O.}~\bibnamefont {{\ifmmode\check{S}\else\v{S}\fi}ipr}},
  \bibinfo {author} {\bibfnamefont {J.}~\bibnamefont {Sinova}},\ and\ \bibinfo
  {author} {\bibfnamefont {T.}~\bibnamefont {Jungwirth}},\ }\bibfield
  {journal} {\bibinfo  {journal} {arXiv}\ }\href
  {https://doi.org/10.48550/arXiv.2211.13806} {10.48550/arXiv.2211.13806}
  (\bibinfo {year} {2022}{\natexlab{c}}),\ \Eprint
  {https://arxiv.org/abs/2211.13806} {2211.13806} \BibitemShut {NoStop}%
\bibitem [{\citenamefont {Fedchenko}\ \emph {et~al.}(2023)\citenamefont
  {Fedchenko}, \citenamefont {Minar}, \citenamefont {Akashdeep}, \citenamefont
  {D'Souza}, \citenamefont {Vasilyev}, \citenamefont {Tkach}, \citenamefont
  {Odenbreit}, \citenamefont {Nguyen}, \citenamefont {Kutnyakhov},
  \citenamefont {Wind}, \citenamefont {Wenthaus}, \citenamefont {Scholz},
  \citenamefont {Rossnagel}, \citenamefont {Hoesch}, \citenamefont
  {Aeschlimann}, \citenamefont {Stadtmueller}, \citenamefont {Klaeui},
  \citenamefont {Schoenhense}, \citenamefont {Jakob}, \citenamefont
  {Jungwirth}, \citenamefont {Smejkal}, \citenamefont {Sinova},\ and\
  \citenamefont {Elmers}}]{RuO2_obser_23}%
  \BibitemOpen
  \bibfield  {author} {\bibinfo {author} {\bibfnamefont {O.}~\bibnamefont
  {Fedchenko}}, \bibinfo {author} {\bibfnamefont {J.}~\bibnamefont {Minar}},
  \bibinfo {author} {\bibfnamefont {A.}~\bibnamefont {Akashdeep}}, \bibinfo
  {author} {\bibfnamefont {S.~W.}\ \bibnamefont {D'Souza}}, \bibinfo {author}
  {\bibfnamefont {D.}~\bibnamefont {Vasilyev}}, \bibinfo {author}
  {\bibfnamefont {O.}~\bibnamefont {Tkach}}, \bibinfo {author} {\bibfnamefont
  {L.}~\bibnamefont {Odenbreit}}, \bibinfo {author} {\bibfnamefont {Q.~L.}\
  \bibnamefont {Nguyen}}, \bibinfo {author} {\bibfnamefont {D.}~\bibnamefont
  {Kutnyakhov}}, \bibinfo {author} {\bibfnamefont {N.}~\bibnamefont {Wind}},
  \bibinfo {author} {\bibfnamefont {L.}~\bibnamefont {Wenthaus}}, \bibinfo
  {author} {\bibfnamefont {M.}~\bibnamefont {Scholz}}, \bibinfo {author}
  {\bibfnamefont {K.}~\bibnamefont {Rossnagel}}, \bibinfo {author}
  {\bibfnamefont {M.}~\bibnamefont {Hoesch}}, \bibinfo {author} {\bibfnamefont
  {M.}~\bibnamefont {Aeschlimann}}, \bibinfo {author} {\bibfnamefont
  {B.}~\bibnamefont {Stadtmueller}}, \bibinfo {author} {\bibfnamefont
  {M.}~\bibnamefont {Klaeui}}, \bibinfo {author} {\bibfnamefont
  {G.}~\bibnamefont {Schoenhense}}, \bibinfo {author} {\bibfnamefont
  {G.}~\bibnamefont {Jakob}}, \bibinfo {author} {\bibfnamefont
  {T.}~\bibnamefont {Jungwirth}}, \bibinfo {author} {\bibfnamefont
  {L.}~\bibnamefont {Smejkal}}, \bibinfo {author} {\bibfnamefont
  {J.}~\bibnamefont {Sinova}},\ and\ \bibinfo {author} {\bibfnamefont {H.~J.}\
  \bibnamefont {Elmers}},\ }\bibfield  {journal} {\bibinfo  {journal} {arXiv}\
  }\href {https://doi.org/10.48550/arXiv.2306.02170}
  {10.48550/arXiv.2306.02170} (\bibinfo {year} {2023}),\ \Eprint
  {https://arxiv.org/abs/2306.02170} {2306.02170} \BibitemShut {NoStop}%
\bibitem [{\citenamefont {Hariki}\ \emph {et~al.}(2023)\citenamefont {Hariki},
  \citenamefont {Yamaguchi}, \citenamefont {Kriegner}, \citenamefont {Edmonds},
  \citenamefont {Wadley}, \citenamefont {Dhesi}, \citenamefont {Springholz},
  \citenamefont {{\ifmmode\check{S}\else\v{S}\fi}mejkal}, \citenamefont
  {V{\ifmmode\acute{y}\else\'{y}\fi}born{\ifmmode\acute{y}\else\'{y}\fi}},
  \citenamefont {Jungwirth},\ and\ \citenamefont
  {Kune{\ifmmode\check{s}\else\v{s}\fi}}}]{Mnte_23}%
  \BibitemOpen
  \bibfield  {author} {\bibinfo {author} {\bibfnamefont {A.}~\bibnamefont
  {Hariki}}, \bibinfo {author} {\bibfnamefont {T.}~\bibnamefont {Yamaguchi}},
  \bibinfo {author} {\bibfnamefont {D.}~\bibnamefont {Kriegner}}, \bibinfo
  {author} {\bibfnamefont {K.~W.}\ \bibnamefont {Edmonds}}, \bibinfo {author}
  {\bibfnamefont {P.}~\bibnamefont {Wadley}}, \bibinfo {author} {\bibfnamefont
  {S.~S.}\ \bibnamefont {Dhesi}}, \bibinfo {author} {\bibfnamefont
  {G.}~\bibnamefont {Springholz}}, \bibinfo {author} {\bibfnamefont
  {L.}~\bibnamefont {{\ifmmode\check{S}\else\v{S}\fi}mejkal}}, \bibinfo
  {author} {\bibfnamefont {K.}~\bibnamefont
  {V{\ifmmode\acute{y}\else\'{y}\fi}born{\ifmmode\acute{y}\else\'{y}\fi}}},
  \bibinfo {author} {\bibfnamefont {T.}~\bibnamefont {Jungwirth}},\ and\
  \bibinfo {author} {\bibfnamefont {J.}~\bibnamefont
  {Kune{\ifmmode\check{s}\else\v{s}\fi}}},\ }\bibfield  {journal} {\bibinfo
  {journal} {arXiv}\ }\href {https://doi.org/10.48550/arXiv.2305.03588}
  {10.48550/arXiv.2305.03588} (\bibinfo {year} {2023}),\ \Eprint
  {https://arxiv.org/abs/2305.03588} {2305.03588} \BibitemShut {NoStop}%
\bibitem [{\citenamefont {Sun}\ \emph {et~al.}(2023)\citenamefont {Sun},
  \citenamefont {Brataas},\ and\ \citenamefont {Linder}}]{Sun2023_andreev}%
  \BibitemOpen
  \bibfield  {author} {\bibinfo {author} {\bibfnamefont {C.}~\bibnamefont
  {Sun}}, \bibinfo {author} {\bibfnamefont {A.}~\bibnamefont {Brataas}},\ and\
  \bibinfo {author} {\bibfnamefont {J.}~\bibnamefont {Linder}},\ }\href
  {https://doi.org/10.1103/PhysRevB.108.054511} {\bibfield  {journal} {\bibinfo
   {journal} {Phys. Rev. B}\ }\textbf {\bibinfo {volume} {108}},\ \bibinfo
  {pages} {054511} (\bibinfo {year} {2023})}\BibitemShut {NoStop}%
\bibitem [{\citenamefont {Ouassou}\ \emph {et~al.}(2023)\citenamefont
  {Ouassou}, \citenamefont {Brataas},\ and\ \citenamefont
  {Linder}}]{ouassou_prl_23}%
  \BibitemOpen
  \bibfield  {author} {\bibinfo {author} {\bibfnamefont {J.~A.}\ \bibnamefont
  {Ouassou}}, \bibinfo {author} {\bibfnamefont {A.}~\bibnamefont {Brataas}},\
  and\ \bibinfo {author} {\bibfnamefont {J.}~\bibnamefont {Linder}},\ }\href
  {https://doi.org/10.1103/PhysRevLett.131.076003} {\bibfield  {journal}
  {\bibinfo  {journal} {Phys. Rev. Lett.}\ }\textbf {\bibinfo {volume} {131}},\
  \bibinfo {pages} {076003} (\bibinfo {year} {2023})}\BibitemShut {NoStop}%
\bibitem [{\citenamefont {Zhang}\ \emph {et~al.}(2023)\citenamefont {Zhang},
  \citenamefont {Hu},\ and\ \citenamefont {Neupert}}]{zhang_arxiv_2023}%
  \BibitemOpen
  \bibfield  {author} {\bibinfo {author} {\bibfnamefont {S.-B.}\ \bibnamefont
  {Zhang}}, \bibinfo {author} {\bibfnamefont {L.-H.}\ \bibnamefont {Hu}},\ and\
  \bibinfo {author} {\bibfnamefont {T.}~\bibnamefont {Neupert}},\ }\bibfield
  {journal} {\bibinfo  {journal} {arXiv}\ }\href
  {https://doi.org/10.48550/arXiv.2302.13185} {10.48550/arXiv.2302.13185}
  (\bibinfo {year} {2023}),\ \Eprint {https://arxiv.org/abs/2302.13185}
  {2302.13185} \BibitemShut {NoStop}%
\bibitem [{\citenamefont {Papaj}(2023)}]{Papaj2023May}%
  \BibitemOpen
  \bibfield  {author} {\bibinfo {author} {\bibfnamefont {M.}~\bibnamefont
  {Papaj}},\ }\bibfield  {journal} {\bibinfo  {journal} {arXiv}\ }\href
  {https://doi.org/10.48550/arXiv.2305.03856} {10.48550/arXiv.2305.03856}
  (\bibinfo {year} {2023}),\ \Eprint {https://arxiv.org/abs/2305.03856}
  {2305.03856} \BibitemShut {NoStop}%
\bibitem [{\citenamefont {Beenakker}\ and\ \citenamefont
  {Vakhtel}(2023)}]{beenakker_prb_23}%
  \BibitemOpen
  \bibfield  {author} {\bibinfo {author} {\bibfnamefont {C.~W.~J.}\
  \bibnamefont {Beenakker}}\ and\ \bibinfo {author} {\bibfnamefont
  {T.}~\bibnamefont {Vakhtel}},\ }\href
  {https://doi.org/10.1103/PhysRevB.108.075425} {\bibfield  {journal} {\bibinfo
   {journal} {Phys. Rev. B}\ }\textbf {\bibinfo {volume} {108}},\ \bibinfo
  {pages} {075425} (\bibinfo {year} {2023})}\BibitemShut {NoStop}%
\bibitem [{\citenamefont {Jin}\ \emph {et~al.}(2023)\citenamefont {Jin},
  \citenamefont {Yang}, \citenamefont {Zeng}, \citenamefont {Cao},\ and\
  \citenamefont {Yan}}]{magnon_magnon_AM}%
  \BibitemOpen
  \bibfield  {author} {\bibinfo {author} {\bibfnamefont {Z.}~\bibnamefont
  {Jin}}, \bibinfo {author} {\bibfnamefont {H.}~\bibnamefont {Yang}}, \bibinfo
  {author} {\bibfnamefont {Z.}~\bibnamefont {Zeng}}, \bibinfo {author}
  {\bibfnamefont {Y.}~\bibnamefont {Cao}},\ and\ \bibinfo {author}
  {\bibfnamefont {P.}~\bibnamefont {Yan}},\ }\bibfield  {journal} {\bibinfo
  {journal} {arXiv}\ }\href {https://doi.org/10.48550/arXiv.2307.00909}
  {10.48550/arXiv.2307.00909} (\bibinfo {year} {2023}),\ \Eprint
  {https://arxiv.org/abs/2307.00909} {2307.00909} \BibitemShut {NoStop}%
\bibitem [{\citenamefont {Ren}\ \emph {et~al.}(2023)\citenamefont {Ren},
  \citenamefont {Chen}, \citenamefont {Zhu}, \citenamefont {Yu}, \citenamefont
  {Zhang}, \citenamefont {Li}, \citenamefont {Liu}, \citenamefont {Li},\ and\
  \citenamefont {Liu}}]{spin_space_group_AM1}%
  \BibitemOpen
  \bibfield  {author} {\bibinfo {author} {\bibfnamefont {J.}~\bibnamefont
  {Ren}}, \bibinfo {author} {\bibfnamefont {X.}~\bibnamefont {Chen}}, \bibinfo
  {author} {\bibfnamefont {Y.}~\bibnamefont {Zhu}}, \bibinfo {author}
  {\bibfnamefont {Y.}~\bibnamefont {Yu}}, \bibinfo {author} {\bibfnamefont
  {A.}~\bibnamefont {Zhang}}, \bibinfo {author} {\bibfnamefont
  {J.}~\bibnamefont {Li}}, \bibinfo {author} {\bibfnamefont {Y.}~\bibnamefont
  {Liu}}, \bibinfo {author} {\bibfnamefont {C.}~\bibnamefont {Li}},\ and\
  \bibinfo {author} {\bibfnamefont {Q.}~\bibnamefont {Liu}},\ }\bibfield
  {journal} {\bibinfo  {journal} {arXiv}\ }\href
  {https://doi.org/10.48550/arXiv.2307.10369} {10.48550/arXiv.2307.10369}
  (\bibinfo {year} {2023}),\ \Eprint {https://arxiv.org/abs/2307.10369}
  {2307.10369} \BibitemShut {NoStop}%
\bibitem [{\citenamefont {Das}\ \emph {et~al.}(2023)\citenamefont {Das},
  \citenamefont {Suri},\ and\ \citenamefont {Soori}}]{das_arxiv_23}%
  \BibitemOpen
  \bibfield  {author} {\bibinfo {author} {\bibfnamefont {S.}~\bibnamefont
  {Das}}, \bibinfo {author} {\bibfnamefont {D.}~\bibnamefont {Suri}},\ and\
  \bibinfo {author} {\bibfnamefont {A.}~\bibnamefont {Soori}},\ }\bibfield
  {journal} {\bibinfo  {journal} {arXiv}\ }\href
  {https://doi.org/10.1088/1361-648X/acea12} {10.1088/1361-648X/acea12}
  (\bibinfo {year} {2023}),\ \Eprint {https://arxiv.org/abs/2305.06680}
  {2305.06680} \BibitemShut {NoStop}%
\bibitem [{\citenamefont {Jiang}\ \emph {et~al.}(2023)\citenamefont {Jiang},
  \citenamefont {Song}, \citenamefont {Zhu}, \citenamefont {Fang},
  \citenamefont {Weng}, \citenamefont {Liu}, \citenamefont {Yang},\ and\
  \citenamefont {Fang}}]{spin_space_group_AM2}%
  \BibitemOpen
  \bibfield  {author} {\bibinfo {author} {\bibfnamefont {Y.}~\bibnamefont
  {Jiang}}, \bibinfo {author} {\bibfnamefont {Z.}~\bibnamefont {Song}},
  \bibinfo {author} {\bibfnamefont {T.}~\bibnamefont {Zhu}}, \bibinfo {author}
  {\bibfnamefont {Z.}~\bibnamefont {Fang}}, \bibinfo {author} {\bibfnamefont
  {H.}~\bibnamefont {Weng}}, \bibinfo {author} {\bibfnamefont {Z.-X.}\
  \bibnamefont {Liu}}, \bibinfo {author} {\bibfnamefont {J.}~\bibnamefont
  {Yang}},\ and\ \bibinfo {author} {\bibfnamefont {C.}~\bibnamefont {Fang}},\
  }\bibfield  {journal} {\bibinfo  {journal} {arXiv}\ }\href
  {https://doi.org/10.48550/arXiv.2307.10371} {10.48550/arXiv.2307.10371}
  (\bibinfo {year} {2023}),\ \Eprint {https://arxiv.org/abs/2307.10371}
  {2307.10371} \BibitemShut {NoStop}%
\bibitem [{\citenamefont {Bai}\ \emph {et~al.}(2023)\citenamefont {Bai},
  \citenamefont {Zhang}, \citenamefont {Zhou}, \citenamefont {Chen},
  \citenamefont {Wan}, \citenamefont {Han}, \citenamefont {Zhu}, \citenamefont
  {Liang}, \citenamefont {Su}, \citenamefont {Han}, \citenamefont {Pan},\ and\
  \citenamefont {Song}}]{bai_prl23}%
  \BibitemOpen
  \bibfield  {author} {\bibinfo {author} {\bibfnamefont {H.}~\bibnamefont
  {Bai}}, \bibinfo {author} {\bibfnamefont {Y.~C.}\ \bibnamefont {Zhang}},
  \bibinfo {author} {\bibfnamefont {Y.~J.}\ \bibnamefont {Zhou}}, \bibinfo
  {author} {\bibfnamefont {P.}~\bibnamefont {Chen}}, \bibinfo {author}
  {\bibfnamefont {C.~H.}\ \bibnamefont {Wan}}, \bibinfo {author} {\bibfnamefont
  {L.}~\bibnamefont {Han}}, \bibinfo {author} {\bibfnamefont {W.~X.}\
  \bibnamefont {Zhu}}, \bibinfo {author} {\bibfnamefont {S.~X.}\ \bibnamefont
  {Liang}}, \bibinfo {author} {\bibfnamefont {Y.~C.}\ \bibnamefont {Su}},
  \bibinfo {author} {\bibfnamefont {X.~F.}\ \bibnamefont {Han}}, \bibinfo
  {author} {\bibfnamefont {F.}~\bibnamefont {Pan}},\ and\ \bibinfo {author}
  {\bibfnamefont {C.}~\bibnamefont {Song}},\ }\href
  {https://doi.org/10.1103/PhysRevLett.130.216701} {\bibfield  {journal}
  {\bibinfo  {journal} {Phys. Rev. Lett.}\ }\textbf {\bibinfo {volume} {130}},\
  \bibinfo {pages} {216701} (\bibinfo {year} {2023})}\BibitemShut {NoStop}%
\bibitem [{\citenamefont {An}\ \emph {et~al.}(2017)\citenamefont {An},
  \citenamefont {Xiao}, \citenamefont {Tu}, \citenamefont {Yu}, \citenamefont
  {Fal{'}ko},\ and\ \citenamefont {Yao}}]{momentum_transfer_r1}%
  \BibitemOpen
  \bibfield  {author} {\bibinfo {author} {\bibfnamefont {X.-T.}\ \bibnamefont
  {An}}, \bibinfo {author} {\bibfnamefont {J.}~\bibnamefont {Xiao}}, \bibinfo
  {author} {\bibfnamefont {M.~W.-Y.}\ \bibnamefont {Tu}}, \bibinfo {author}
  {\bibfnamefont {H.}~\bibnamefont {Yu}}, \bibinfo {author} {\bibfnamefont
  {V.~I.}\ \bibnamefont {Fal{'}ko}},\ and\ \bibinfo {author} {\bibfnamefont
  {W.}~\bibnamefont {Yao}},\ }\href
  {https://doi.org/10.1103/PhysRevLett.118.096602} {\bibfield  {journal}
  {\bibinfo  {journal} {Phys. Rev. Lett.}\ }\textbf {\bibinfo {volume} {118}},\
  \bibinfo {pages} {096602} (\bibinfo {year} {2017})}\BibitemShut {NoStop}%
\bibitem [{\citenamefont {Guan}\ \emph {et~al.}(2020)\citenamefont {Guan},
  \citenamefont {Zhang}, \citenamefont {Wang}, \citenamefont {Yu},
  \citenamefont {Xia},\ and\ \citenamefont {Li}}]{momentum_transfer_2}%
  \BibitemOpen
  \bibfield  {author} {\bibinfo {author} {\bibfnamefont {J.-H.}\ \bibnamefont
  {Guan}}, \bibinfo {author} {\bibfnamefont {Y.-Y.}\ \bibnamefont {Zhang}},
  \bibinfo {author} {\bibfnamefont {S.-S.}\ \bibnamefont {Wang}}, \bibinfo
  {author} {\bibfnamefont {Y.}~\bibnamefont {Yu}}, \bibinfo {author}
  {\bibfnamefont {Y.}~\bibnamefont {Xia}},\ and\ \bibinfo {author}
  {\bibfnamefont {S.-S.}\ \bibnamefont {Li}},\ }\href
  {https://doi.org/10.1103/PhysRevB.102.064203} {\bibfield  {journal} {\bibinfo
   {journal} {Phys. Rev. B}\ }\textbf {\bibinfo {volume} {102}},\ \bibinfo
  {pages} {064203} (\bibinfo {year} {2020})}\BibitemShut {NoStop}%
\bibitem [{\citenamefont {Blonder}\ and\ \citenamefont
  {Tinkham}(1983)}]{mismatch_barrier_1}%
  \BibitemOpen
  \bibfield  {author} {\bibinfo {author} {\bibfnamefont {G.~E.}\ \bibnamefont
  {Blonder}}\ and\ \bibinfo {author} {\bibfnamefont {M.}~\bibnamefont
  {Tinkham}},\ }\href {https://doi.org/10.1103/PhysRevB.27.112} {\bibfield
  {journal} {\bibinfo  {journal} {Phys. Rev. B}\ }\textbf {\bibinfo {volume}
  {27}},\ \bibinfo {pages} {112} (\bibinfo {year} {1983})}\BibitemShut
  {NoStop}%
\end{thebibliography}%

\appendix

\begin{widetext}

\section{Expressions in the AM1}
The effective low-energy Hamiltonian for the AM1, as shown in Fig. 1(a) in the main text, using an electron field operator basis $\psi = [\psi_\uparrow, \psi_\downarrow]^T$, is given by
\begin{equation}
{H}_\text{AM} =  -{\frac{\hbar^2\triangledown^2}{2 m_e}} - \mu + \alpha{\sigma_z}k_x k_y,
\end{equation}
in which $\alpha$ is the parameter characterizes the altermagnetism strength, ${\sigma_z}$ denotes the Pauli matrix, $m_e$ is the electron mass and $\mu$ is the chemical potential. The two eigenpairs are obtained as: $E_1=E_+$ with $(1,0)^T$ for $e\uparrow$ (electron with spin-up) and $E_2=E_-$ with $(0,1)^T$ for $e\downarrow$ (electron with spin-down). The eigenenergies are described by  
\begin{equation}
E_\pm=\frac{\hbar^2(k_{x}^2+k_{y}^2)}{2m_e}-\mu \pm \alpha k_x k_y.
\label{eq:Epm}
\end{equation}

Applying $E_1=E_2=E$, the $x$-components of the wave vectors in the AM are given by
\begin{equation}
   k_{e\uparrow,\pm}=\pm\frac{1}{\hbar}\sqrt{2m_e(\mu+E)-\hbar^2 k_y^2 +\frac{\alpha^2 m_e^2 k_y^2}{\hbar^2}}-\frac{\alpha m_e k_y}{\hbar^2}, 
\label{eq:keup}
\end{equation}
\begin{equation}
k_{e\downarrow,\pm}=\pm\frac{1}{\hbar}\sqrt{2m_e(\mu+E)-\hbar^2 k_y^2 +\frac{\alpha^2 m_e^2 k_y^2}{\hbar^2}}+\frac{\alpha m_e k_y}{\hbar^2}, 
\label{eq:kedn}
\end{equation}
 in which the $\pm$ sign in the subscript denotes the propagation direction along the $\pm x$. Here we assume translational invariance in the $y$-direction with belonging conserved momentum $k_y$. The momentum $k_y$ of the incident particle appearing in Eqs. (\ref{eq:keup},\ref{eq:kedn}) is determined by the Fermi surface of the incident particle, which is described as follows.

 Consider an $e\uparrow$ particle in the AM. We then have $E = E_+ =\frac{\hbar^2(k_{x}^2+k_{y}^2)}{2m_e}-\mu + \alpha k_x k_y$ in Eq. (\ref{eq:Epm}), which defines an elliptical Fermi surface in the $\boldsymbol{k}$-space when $\alpha<\hbar^2/m_e\equiv\alpha_c$.
On the other hand, Eq. (\ref{eq:Epm}) corresponds to a hyperbola when $\alpha > \alpha_c$, which can not define a closed integral path. Therefore, we confine $\alpha < \alpha_c$ in this work. The general equation of the ellipse is given by
\begin{equation}
\frac{\hbar^2k_{x}^2}{2m_e}+\alpha k_x k_y + \frac{\hbar^2 k_{y}^2}{2m_e}  - (\mu + E) = 0,
\end{equation}
from which the semi-major (minor) axis can be obtained as
\begin{equation}
    a_1 = \sqrt{\frac{2m_e(\mu+E)}{\hbar^2-m_e \alpha}},\quad
    b_1 = \sqrt{\frac{2m_e(\mu+E)}{\hbar^2+m_e\alpha}}.
\end{equation}
Consequently, the wave vectors on the Fermi surface of $e\uparrow$ in the AM are described by
\begin{equation}
k_{y,e\uparrow} = r_1 \sin\theta
,\quad k_{x,e\uparrow} = r_1 \cos\theta,\quad r_1 = \frac{a_1 b_1}{\sqrt{b_1^2 \cos^2(\theta+\pi/4)+a_1^2 \sin^2(\theta+\pi/4)}},
\label{eq:ky_eup}
\end{equation}
in which $\theta$ is the incident angle in the AM with respect to the $x$-axis. %Therefore, we use $k_y = k_{y,e\uparrow}$ in Eqs. (\ref{eq:keup},\ref{eq:kedn}) to get the $x$-component of the wave vector belonging to incident $e\uparrow$ particles on the AM1 side.

Similarly, we can obtain the wave vectors on the Fermi surface of $e\downarrow$ particle in the AM, i.e.,
\begin{align}
k_{y,e\downarrow} &= r_2 \sin\theta
,\quad k_{x,e\downarrow} = r_2 \cos\theta,\notag \\ r_2 &= \frac{a_2 b_2}{\sqrt{b_2^2 \cos^2(\theta-\pi/4)+a_2^2 \sin^2(\theta-\pi/4)}},\quad   a_2 = \sqrt{\frac{2m_e(\mu+E)}{\hbar^2-m_e \alpha}},\quad  b_2= \sqrt{\frac{2m_e(\mu+E)}{\hbar^2+m_e\alpha}}. \label{eq:ky_edn}
\end{align}
%By inserting $k_y=k_{y,e\downarrow}$ into Eqs. (\ref{eq:keup},\ref{eq:kedn}) we can get the  $x$-components of the wave vectors induced by $e\downarrow$ incident on the AM1 side, which will appear in the wave functions to describe the propagation along the $x$-direction. 

% Note the relation between the two $x$-components of the wave vectors involved, e.g., $k_{x,e\uparrow}$ and $k_{e\uparrow,\pm}$: $k_{x,e\uparrow}$ is the $x$-component of the wave vector of $e\uparrow$ particle on the Fermi surface for a given value of the angle $\theta$, and is thus uniquely defined. Instead, $k_{e\uparrow,\pm}$ are the two possible solutions for the $x$-component of the momentum on the Fermi surface which both have the same value for $k_y$. Thus, $k_{e\uparrow,\pm}$ can be used to describe the $x$-component of incident and reflected $e\uparrow$ particles for a given $k_y$-value. Only when considering the $e\uparrow$ incident from the AM with $k_y=k_{y,e\uparrow}$, $k_{x,e\uparrow}$ is equivalent to either $k_{e\uparrow,+}$ or $k_{e\uparrow,-}$ depending on the value of $\theta$. To construct the wave functions, we have thus assumed $k_y$ invariance and include the $x$-components of the wave vectors for different scattered particles to describe the reflection and transmission procsesses. In effect, Eqs. (\ref{eq:keup},\ref{eq:kedn}) are utilized as wave vectors in the wave functions.

Consider the $e\uparrow$ incident from the AM side based on the FI/AM bilayer, we have
\begin{equation}
\Psi_{\text{AM},e\uparrow}=
\left[\begin{pmatrix}
1\\0
\end{pmatrix}e^{ik_{e\uparrow,-}x}+r
\begin{pmatrix}
 1\\0   
\end{pmatrix}e^{ik_{e\uparrow,+}x}\right]e^{-\frac{iEt}{\hbar}} + r^{'}
\begin{pmatrix}
    0\\1
\end{pmatrix}e^{ik^{'}_{e\downarrow,+}x}e^{-\frac{iE^{'}t}{\hbar}}, 
\label{eq:AM_e_up}
\end{equation}
in which we use $k_y = k_{y,e\uparrow}$ given in Eq. (\ref{eq:ky_eup}). $r$ and $r^{'}$ are coefficients describing reflection without and with spin-flip, respectively. These coefficients can be determined by applying appropriate boundary conditions, which we will get back to. To differentiate it from the incident energy $E$, the energy after the spin-flip is denoted as $E^{'}$. Similarly, $k_{e\uparrow,\pm}^{'}$ and $k_{e\downarrow,\pm}^{'}$ have the same forms as shown in Eqs. (\ref{eq:keup},\ref{eq:kedn}) with respect to $E^{'}$.

Consider the $e\downarrow$ incident from the AM side based on the FI/AM bilayer, we have
\begin{equation}
\Psi_{\text{AM},e\downarrow}=
\left[\begin{pmatrix}
0\\1
\end{pmatrix}e^{ik_{e\downarrow,-}x}+r
\begin{pmatrix}
 0\\1   
\end{pmatrix}e^{ik_{e\downarrow,+}x}\right]e^{-\frac{iEt}{\hbar}} + r^{'}
\begin{pmatrix}
    1\\0
\end{pmatrix}e^{ik^{'}_{e\uparrow,+}x}e^{-\frac{iE^{'}t}{\hbar}}, 
\label{eq:AM_e_up}
\end{equation}
in which we use $k_y = k_{y,e\downarrow}$ given in Eq. (\ref{eq:ky_edn}).

\section{Expressions in the FI}
In the FI, the Hamiltonian for electron-like quasiparticles has the form
\begin{equation}
    {H}_{\text{FI}} =  -{\frac{\hbar^2\triangledown^2}{2 m_e}} + U + J\hat{\boldsymbol{\sigma}}\cdot \boldsymbol{M}(t),
\end{equation}
in which $\hat{\boldsymbol{\sigma}}$ denotes the Pauli matrix vector. The potential $U$ is larger than the chemical potential $\mu$ in the nearby AM. $J$ decribes the exchange interaction in the ferromagnet between the localized spin magnetization and the itinerant electrons. The normalized magnetization is defined as 
\begin{equation}
   \boldsymbol{M}(t) = (m \cos\Omega t, m \sin \Omega t, \sqrt{1-m^2}),
\end{equation}
where $m \in [0,1]$ is the magnetization oscillation amplitude and $\Omega$ denotes the FMR frequency for spin pumping. By employing a wavefunction with the structure $ e^{-\frac{iEt}{\hbar}}(e^{-\frac{i \Omega t}{2}},e^{\frac{i \Omega t}{2}})^T $ for its time-dependence, the non-stationary Schr\"odinger equation can be solved as an eigenvalue problem. The two eigenpairs are obtained as: $E_1=E_+$ with $(a_+,b_+)^T$ and $E_2=E_-$ with $(a_-,b_-)^T$. In terms of the adiabaticity parameter $\beta=\frac{\hbar\Omega}{2J}$, the eigenenergies are given by  
\begin{equation}
E_\pm=U+\frac{\hbar^2(k_{x}^2+k_{y}^2)}{2m_e}\pm J\sqrt{1-2\beta\sqrt{1-m^2}+\beta^2}.
\end{equation} 
The corresponding eigenstates are described by the coefficients
\begin{align}
    a_\pm &= \frac{\eta_\pm}{\sqrt{\eta_\pm^2+1}}, \qquad b_\pm= \frac{1}{\sqrt{\eta_\pm^2+1}},\\
    \eta_\pm &= \frac{\sqrt{1-m^2}-\beta \pm \sqrt{1-2\beta\sqrt{1-m^2}+\beta^2}}{m},
\end{align}
which satisfy
\begin{equation}
    a_-=-b_+, \qquad b_-=a_+.
\label{eq:ab}
\end{equation}
Note when $m=0$, we have $a_+ =1$, $b_+=0$, $a_-=0$ and $b_-=1$.

Based on the above, the total wavefunction in the FI is constructed as 
\begin{equation}
\Psi_\text{FI}=t
\begin{pmatrix}
 a_+e^{\frac{-i\Omega t}{2}}\\b_+e^{\frac{i\Omega t}{2}}  
\end{pmatrix}e^{-ik_{\text{F}1}x}e^{\frac{-iE_{1}t}{\hbar}} + p\begin{pmatrix}
    a_-e^{\frac{-i\Omega t}{2}}\\b_-e^{\frac{i\Omega t}{2}}
\end{pmatrix}e^{-ik_{\text{F}2}x}e^{\frac{-iE_{2}t}{\hbar}},
\end{equation}
where $t$ and $p$ are transmission coefficients to be determined by applying appropriate boundary conditions. 

Consider the $e\uparrow$ incident from the AM side based on the FI/AM bilayer, in order to match the time-dependence of the wavefunction components on the FI and AM sides, we can obtain $E^{'}=E-\hbar\Omega$ and $E_1=E_2=E-\frac{\hbar\Omega}{2}$. In terms of $E$, the corresponding $x$ component of the two wave vectors in the FI are expressed as 
\begin{equation}k_{\text{F}1,e\uparrow}=\frac{\sqrt{2m_e[E-U-J(\sqrt{1-2\beta\sqrt{1-m^2}+\beta^2}+\beta)]-\hbar^2k_y^2}}{\hbar},
\label{eq:kF1_eup}
\end{equation}
\begin{equation}k_{\text{F}2,e\uparrow}=\frac{\sqrt{2m_e[E-U+J(\sqrt{1-2\beta\sqrt{1-m^2}+\beta^2}}-\beta)]-\hbar^2k_y^2}{\hbar}.
\label{eq:kF2_eup}
\end{equation} 
Based on the above, we write down
\begin{equation}
\Psi_{\text{FI},e\uparrow}=t
\begin{pmatrix}
 a_+e^{\frac{-i\Omega t}{2}}\\b_+e^{\frac{i\Omega t}{2}}  
\end{pmatrix}e^{-ik_{\text{F}1,e\uparrow}x}e^{\frac{-iE_1t}{\hbar}} + p\begin{pmatrix}
    a_-e^{\frac{-i\Omega t}{2}}\\b_-e^{\frac{i\Omega t}{2}}
\end{pmatrix}e^{-ik_{\text{F}2,e\uparrow}x}e^{\frac{-iE_2t}{\hbar}}.
\end{equation}

Consider the $e\downarrow$ incident from the AM side based on the FI/AM bilayer, in order to match the time-dependence of the wavefunction components on the FI and AM sides, we can obtain $E^{'}=E+\hbar\Omega$ and $E_1=E_2=E+\frac{\hbar\Omega}{2}$. In terms of $E$, the corresponding $x$ component of the two wave vectors in the FI are expressed as 
\begin{equation}
k_{\text{FI}1,e\downarrow}=\frac{\sqrt{2m_e[E-U-J(\sqrt{1-2\beta\sqrt{1-m^2}+\beta^2}-\beta)]-\hbar^2k_y^2}}{\hbar},
\end{equation}
\begin{equation}
k_{\text{F}2,e\downarrow}=\frac{\sqrt{2m_e[E-U+J(\sqrt{1-2\beta\sqrt{1-m^2}+\beta^2}+ \beta)]-\hbar^2k_y^2}}{\hbar}.\end{equation} 
Based on the above, we write down
\begin{equation}
\Psi_{\text{FI},e\downarrow}=t
\begin{pmatrix}
 a_+e^{\frac{-i\Omega t}{2}}\\b_+e^{\frac{i\Omega t}{2}}  
\end{pmatrix}e^{-ik_{\text{F}1,e\downarrow}x}e^{\frac{-iE_1t}{\hbar}} + p\begin{pmatrix}
    a_-e^{\frac{-i\Omega t}{2}}\\b_-e^{\frac{i\Omega t}{2}}
\end{pmatrix}e^{-ik_{\text{F}2,e\downarrow}x}e^{\frac{-iE_2t}{\hbar}}.
\end{equation}

Note that all wave numbers in the FI possess imaginary values since a large potential $U$ is required to ensure the ferromagnet to be insulating. To ensure this, we use $U=2\mu$ throughout this work.

\section{Wavefunctions in the AM and FI}
Here we summarize the wavefunctions in the AM and FI, in which the time-dependence is omitted since we have applied equal time-dependence on both sides.

Consider the $e\uparrow$ incident from the AM side based on the FI/AM bilayer, we have
\begin{equation}
\Psi_{\text{AM},e\uparrow}=
\begin{pmatrix}
1\\0
\end{pmatrix}e^{ik_{e\uparrow,-}x}+r
\begin{pmatrix}
 1\\0   
\end{pmatrix}e^{ik_{e\uparrow,+}x}+ r^{'}
\begin{pmatrix}
    0\\1
\end{pmatrix}e^{ik^{'}_{e\downarrow,+}x}, 
\label{eq:AM_e_up1}
\end{equation}
\begin{equation}
\Psi_{\text{FI},e\uparrow}=t
\begin{pmatrix}
 a_+\\b_+  
\end{pmatrix}e^{-ik_{\text{F}1,e\uparrow}x} + p\begin{pmatrix}
    a_-\\b_-
\end{pmatrix}e^{-ik_{\text{F}2,e\uparrow}x},
\end{equation}
in which $E^{'}=E-\hbar\Omega$.

Consider the $e\downarrow$ incident from the AM side based on the FI/AM bilayer, we have
\begin{equation}
\Psi_{\text{AM},e\downarrow}=
\begin{pmatrix}
0\\1
\end{pmatrix}e^{ik_{e\downarrow,-}x}+r
\begin{pmatrix}
 0\\1   
\end{pmatrix}e^{ik_{e\downarrow,+}x} + r^{'}
\begin{pmatrix}
    1\\0
\end{pmatrix}e^{ik^{'}_{e\uparrow,+}x}, 
\label{eq:AM_e_up1}
\end{equation}
\begin{equation}
\Psi_{\text{FI},e\downarrow}=t
\begin{pmatrix}
 a_+\\b_+ 
\end{pmatrix}e^{-ik_{\text{F}1,e\downarrow}x} + p\begin{pmatrix}
    a_-\\b_-
\end{pmatrix}e^{-ik_{\text{F}2,e\downarrow}x},
\end{equation}
in which $E^{'}=E+\hbar\Omega$.

\section{Boundary conditions}
\label{sec:BC}
We consider a planar FI in contact with AM1 through a Rashba spin-orbit coupled interface. This interfacial contribution to the Hamiltonian takes the form
\begin{align}
H_I &= [U_0 + \frac{U_\text{SO}}{k_F} \hat{\boldsymbol{n}} \cdot (\hat{\boldsymbol{\sigma} }\times \boldsymbol{k}) ]\delta(x)\notag\\
&=[U_0 - \frac{U_\text{SO}}{k_F}k_y\sigma_z]\delta(x),
\end{align}
in which we take $\boldsymbol{n} = \boldsymbol{x}$ and $k_F=\sqrt{2m_e\mu}/\hbar$ is the Fermi wave vector.
This $\delta$-function will influence the boundary conditions that the scattering wavefunctions have to satisfy. Consequently, the Hamiltonian of the bilayer system becomes
\begin{equation}
    {H} =  -\frac{\hbar^2\nabla^2}{2m_e} + H_I + \frac{\alpha k_y}{2} \{k_x,\Theta(x)\} {\sigma_z},
\label{eq:BC}
\end{equation}
in which only the terms affecting the boundary conditions are included. Note here antisymmetrization of the altermagnetic term $\alpha k_xk_y{\sigma_z} \to \frac{\alpha k_y}{2} \{k_x,\Theta(x)\} {\sigma_z}$ is necessary to ensure hermiticity of the Hamilton-operator, where $\Theta(x)$ is the step function. Above, $k_x = -\i\partial_x$.

Eq. (\ref{eq:BC}) can be rewritten as 
\begin{equation}
    {H} =  -\frac{\hbar^2\nabla^2}{2m_e} + (U_0-\frac{U_\text{SO}}{k_F}k_y \sigma) \delta(x) + \frac{\alpha k_y \sigma}{2} \{k_x,\Theta(x)\},
    \label{eq:BC2}
\end{equation}
where $\sigma=+1(-1)$ for $e\uparrow (\downarrow)$. In Eq. (\ref{eq:BC2}), we have
\begin{equation}
\begin{aligned}
    \{k_x,\Theta(x)\}\Psi &= k_x[\Theta(x)\Psi]+\Theta(x)(k_x\Psi)\\
    &=-\i[\Psi\partial_x\Theta(x)+\Theta(x)\partial_x \Psi]-\i\Theta(x)\partial_x\Psi\\
    &=-\i\delta(x)\Psi-2\i\Theta(x)\partial_x\Psi.
\end{aligned}
\end{equation}

Apply $H\Psi = E \Psi$ and integrate over $[-\epsilon,\epsilon]$ with $\epsilon \rightarrow 0$, we have
\begin{align}
    \int_{-\epsilon}^{+\epsilon}\partial_x^2\Psi dx &= \frac{2m_e}{\hbar^2}\int_{-\epsilon}^{+\epsilon}[U_0 - (\frac{\i \alpha}{2} + \frac{U_\text{SO}}{k_F})k_y\sigma]\delta(x)\Psi dx \notag\\
    &- \frac{2m_e}{\hbar^2}\int_{-\epsilon}^{+\epsilon}\i\alpha k_y \sigma \Theta(x) \partial_x \Psi dx - \frac{2m_e}{\hbar^2}\int_{-\epsilon}^{+\epsilon} E\Psi dx.
\end{align}
Consequently, the remaining nonzero terms are
\begin{equation}
\partial_x\Psi\big|_{+\epsilon}-\partial_x\Psi\big|_{-\epsilon}=\frac{2m_e}{\hbar^2}[U_0 - (\frac{\i\alpha}{2}+\frac{U_\text{SO}}{k_F})k_y\sigma]\Psi\big|_{+\epsilon}
\end{equation}
with $\Psi\big|_{+\epsilon} = \Psi\big|_{-\epsilon}$ and $\sigma=+1(-1)$ for $e\uparrow (\downarrow)$.

For notation convenience, we rewrite the boundary conditions for $e\uparrow$ incident from the AM side based on the FI/AM bilayer as
\begin{equation}
\Psi_{\text{AM}}\big|_{x=0}=\Psi_{\text{FI}}\big|_{x=0}=\begin{pmatrix}
    f\\g
\end{pmatrix},
\label{eq:BC_1}
\end{equation} 
\begin{equation}
    \partial_x\Psi_{\text{AM}}\big|_{x=0}-\partial_x\Psi_{\text{FI}}\big|_{x=0}=\begin{pmatrix}
k_{\alpha,+1}f\\k_{\alpha,-1}g
\end{pmatrix},
\label{eq:BC_2}
\end{equation} 
where $k_{\alpha,\sigma} =\frac{2m_e}{\hbar^2}[U_0-(\frac{\i\alpha}{2}+\frac{U_\text{SO}}{k_F})k_y\sigma] $ with $\sigma=+1(-1)$.

The boundary conditions for $e\downarrow$ incident from the AM side have the same forms as Eqs. (\ref{eq:BC_1},\ref{eq:BC_2}) with different explicit expressions for $f$ and $g$.

\section{1D DOS and 2D DOS in the AM}
\label{sec:DOS}
For $e\uparrow$ incident from the AM1 side based on the FI/AM1 bilayer, we have 
\begin{equation}
E = E_+ =\frac{\hbar^2(k_{x}^2+k_{y}^2)}{2m_e}-\mu + \alpha k_x k_y, 
\label{eq:ellipse}
\end{equation}
based on which the 1D density of states (DOS) can be calculated as
\begin{equation}
    dk_x/dE=(\frac{\hbar^2k_x}{m_e}+\alpha k_y)^{-1}.
\end{equation}
On the other hand, the general expression for 2D DOS is given by
\begin{equation}
   N(E) = \frac{1}{4\pi^2} \int \frac{dl}{|\nabla_{\veck} E(\veck)|}, 
\label{eq:def_dos}
\end{equation}
which can be used for anisotropic DOS.
%i) When $\alpha < \hbar^2/m_e$, Eq. (\ref{eq:ellipse}) defines an elliptical energy surface. 
In Eq. (\ref{eq:def_dos}), we can use
\begin{equation}
    dl = \sqrt{(\frac{dk_x}{d\theta})^2+(\frac{dk_y}{d\theta})^2}d\theta,
\label{eq:dl}
\end{equation}
\begin{align}
    |\nabla_{\veck} E(\veck)| &= \sqrt{(\frac{\partial E}{\partial k_x})^2 + (\frac{\partial E}{\partial k_y})^2}\notag \\
    &= \sqrt{(\frac{\hbar^2 k_x}{m_e} + \alpha k_y)^2 + (\frac{\hbar^2 k_y}{m_e} + \alpha k_x)^2}.
\label{eq:dEdk}
\end{align}
Insert $k_x = k_{x,e\uparrow}$ and $k_y = k_{y,e\uparrow}$ in Eq. (\ref{eq:ky_eup}) into Eqs. (\ref{eq:dl}) and (\ref{eq:dEdk}), $|\nabla_{\veck} E(\veck)|$ is expressed in terms of $E$ and $\theta$, i.e., $|\nabla_{\veck} E(\veck)| = K(E,\theta)$. Consequently, Eq. (\ref{eq:def_dos}) can be rewritten as 
\begin{align}
    N(E) &= \int_0^{2\pi} N(E,\theta) d\theta,\\
    N(E,\theta) &= \frac{1}{4\pi^2} \frac{\sqrt{(dk_{x,e\uparrow}/d\theta)^2+(dk_{y,e\uparrow}/d\theta)^2}}{K(E,\theta)}
\label{eq:DOS_AM_eup_+}
\end{align}
in which $N(E,\theta)$ corresponds to the DOS at a given incident angle $\theta$. 

%ii) When $\alpha > \hbar^2/m_e$, Eq. (\ref{eq:ellipse}) corresponds to a hyperbola, which can not define a closed integral path. Therefore, we confine $\alpha < \hbar^2/m_e$ in this work, as mentioned before. 

Following the same procedure as described above, the DOS in the AM for $e\downarrow$ incident with $E=E_-$ can be calculated. Note that 1D DOS instead of 2D DOS is utilized in the main text since $\int k_y$ is included separately.

\section{Backflow spin-current}
%The longitudinal spin-pumping current polarized along the $z$ axis can be calculated as 
%\begin{equation}
    %I_{s,\Chi{e\uparrow(\downarrow)}}=\int\Chi{\frac{dk_{y,e\uparrow(\downarrow)}}{d\theta}d\theta}\int j_{sz,\Chi{e\uparrow(\downarrow)}}(E,\theta)f_0(E)\Chi{N_{\text{AM},\Chi{e\uparrow(\downarrow)}}(E,\theta)}dE,
%\label{eq:Is}
%\end{equation}
%where the quantum mechanical spin current $j_{sz,\Chi{e\uparrow(\downarrow)}}(E,\theta)=\Chi{\frac{\hbar^2}{2m_e}(\Imag\{f^*\nabla f\} - \Imag\{g^*\nabla g\} )+\frac{\alpha k_y}{2}(|f|^2+|g|^2)}$ produced by \Chi{$e\uparrow(\downarrow)$ incident from the AM}. $f_0(E)$ denotes the Fermi-Dirac distribution and \Chi{$N_{\text{AM},e\uparrow(\downarrow)}(E,\theta)$} is the density of states in the AM \Chi{for $e\uparrow(\downarrow)$}. \Chi{The total spin-pumping current is calculated as $I_s = I_{s,e\uparrow} + I_{s,e\downarrow}$.} 
In general, a backflow spin current exists due to a spin accumulation that is built up in the material connected to the precessing FI \cite{tserkovnyak_rmp_04}. This backflow current diminishes the magnitude of the total spin current flowing across the interface. Assuming that the material which the spin current is pumped into act as a highly conductive reservoir which drains the spin current, the backflow spin current may be neglected. For a ferromagnet/normal metal bilayer with a Rashba spin-orbit coupled interface, as in the present system, Ref. \cite{SP_SOC_Shufeng} derived a backflow factor $\xi \propto (\lambda_\text{sd}/l_\text{mfp}) \text{coth}(d_N/\lambda_\text{sd})$ where $\lambda_\text{sd}$ is the spin diffusion length, $l_\text{mfp}$ is the electronic mean free path, and $d_N$ is the thickness of the normal layer. For ballistic, large reservoirs, $\xi \to 0$.

\section{45-degree rotated AM2}
Here we summarize the useful equations for the rotated Hamiltonian of AM2 shown in Fig. 1(b) in the main text, i.e.,
\begin{equation}
    {H}_\text{AM} =  -{\frac{\hbar^2\triangledown^2}{2 m_e}} - \mu + \frac{\alpha}{2}(k_x^2- k_y^2){\sigma_z},
\end{equation}
which corresponds to a 45 degree
rotation of the FI/AM1 interface. 
\begin{itemize}
\item 1: eigenpairs:
 
 The two eigenpairs are obtained as: $E_1=E_+$ with $(1,0)^T$ for $e\uparrow$ and  $E_2=E_-$ with $(0,1)^T$ for $e\downarrow$. The eigen-energies are described by  
\begin{equation}
E_\pm=\frac{\hbar^2(k_{x}^2+k_{y}^2)}{2m_e}-\mu \pm \frac{\alpha}{2} (k_x^2-k_y^2).
\end{equation} 

\item 2: wave vectors in the AM to construct the wave functions:
\begin{align}
k_{e\uparrow,\pm}&=\pm \sqrt{\frac{2m_e(\mu+E+\alpha k_y^2/2)-\hbar^2k_y^2}{\hbar^2+m_e \alpha}},\\
k_{e\downarrow,\pm}&=\pm \sqrt{\frac{2m_e(\mu+E-\alpha k_y^2/2)-\hbar^2k_y^2}{\hbar^2-m_e \alpha}}.
\end{align}

\item 3: wave vectors on the AM Fermi surface:
\begin{align}
k_{y,e\uparrow} &=  r_1 \sin\theta
,\quad k_{x,e\uparrow} = r_1 \cos\theta,\notag \\r_1 &= \frac{a_1 b_1}{\sqrt{b_1^2 \cos^2(\theta+\pi/2)+a_1^2 \sin^2(\theta+\pi/2)}},\quad a_1 = \sqrt{\frac{2m_e(\mu+E)}{\hbar^2-m_e \alpha}},\quad b_1= \sqrt{\frac{2m_e(\mu+E)}{\hbar^2+m_e\alpha}}\\
k_{y,e\downarrow} &= r_2 \sin\theta
,\quad k_{x,e\downarrow} = r_2 \cos\theta,\notag \\r_2 &= \frac{a_2 b_2}{\sqrt{b_2^2 \cos^2\theta+a_2^2 \sin^2\theta}},\quad a_2 = \sqrt{\frac{2m_e(\mu+E)}{\hbar^2-m_e \alpha}},\quad b_2= \sqrt{\frac{2m_e(\mu+E)}{\hbar^2+m_e\alpha}}.
\end{align}
%Following the same approach as described in Sec. \ref{sec:DOS}, the DOS can be derived based on the above new wavevectors, e.g., the DOS at angle $\theta$ for $e\uparrow$ incident from the AM is calculated as
%\begin{equation}
    %N(E,\theta) &= \frac{1}{4\pi^2} \frac{\sqrt{(dk_{x,e\uparrow}/d\theta)^2+(dk_{y,e\uparrow}/d\theta)^2}}{\sqrt{(\hbar^2/m_e+\alpha)^2k_{x,e\uparrow}^2+(\hbar^2/m_e-\alpha)^2k_{y,e\uparrow}^2}}.
%\end{equation}

\item 4: boundary conditions:
\begin{equation}
\Psi_{\text{AM}}\big|_{x=0}=\Psi_{\text{FI}}\big|_{x=0}=\begin{pmatrix}
    f\\g
\end{pmatrix},
\end{equation}
\begin{equation}
    \begin{pmatrix}
(1+m_e\alpha/\hbar^2)\partial_x f_\text{AM}\\(1-m_e\alpha/\hbar^2)\partial_x g_\text{AM}
\end{pmatrix}\big|_{x=0} -\partial_x\Psi_{\text{FI}}\big|_{x=0}=\begin{pmatrix}
k_{\alpha,+1}f\\k_{\alpha,-1}g
\end{pmatrix}
\end{equation} 
for $e\uparrow$ and $e\downarrow$ incidents, in which $k_{\alpha,\sigma} =\frac{2m_e}{\hbar^2}[U_0-\frac{U_\text{SO}}{k_F}k_y\sigma] $ with $\sigma=+1(-1)$. To get the above boundary conditions, we follow the similar procedure as described in Sec. \ref{sec:BC} by considering the Hermitian Hamiltonian of the bilayer system as 
\begin{equation}
    {H} =  -\frac{\hbar^2\nabla^2}{2m_e} + H_I + \frac{\alpha}{2}[k_x \Theta(x) k_x - k_y \Theta(x) k_y]{\sigma_z},
\end{equation}
in which $k_x = -\i\partial_x$.

\item 5: The longitudinal quantum mechanical spin current polarized along the $z$ axis in the AM:
\begin{equation}
j_{sz,e\uparrow(\downarrow)}=\frac{\hbar^2}{2m_e}(\Imag\{f^*\nabla f\} - \Imag\{g^*\nabla g\} )+\frac{\alpha}{2}(\Imag\{f^*\nabla f\} + \Imag\{g^*\nabla g\}).
\end{equation}

\end{itemize}

\section{Spin-flip probability}
Here we consider FI/AM1 as an example. If we write the wave function in the form of $\Psi = (f,g)^T$, the probability current in the AM is given by 
\begin{equation}
j_P^{\text{AM}} = \frac{\hbar}{m_e} [\Imag{\{f^*\nabla f \}} + \Imag{\{g^*\nabla g \}}] + \frac{\alpha k_y}{\hbar}(|f|^2-|g|^2).
\label{eq:jP_AM_0}
\end{equation}

Consider the $e\uparrow$ incident from the AM side based on the FI/AM bilayer, we have $\Psi_{\text{AM},e\uparrow}$ with $f = e^{i k_{e\uparrow,-}x} + r e^{ik_{e\uparrow,+}x} $ and $g = r^{'} e^{ik_{e\downarrow,+}^{'}x}$, we have
\begin{equation}
\begin{aligned}
  \Imag{\{f^* \nabla f\}} &= k_{e\uparrow,-}e^{-2\Imag[k_{e\uparrow,-}]x}+ k_{e\uparrow,+}|r|^2e^{-2\Imag[k_{e\uparrow,+}]x}+\Real[(k_{e\uparrow,+}+k_{e\uparrow,-}^{*})re^{i(k_{e\uparrow,+}-k_{e\uparrow,-}^{*})x}],\\
  \Imag{\{g^* \nabla g\}} &= k_{e\downarrow,+}^{'}|r^{'}|^2 e^{-2\Imag[k_{e\downarrow,+}^{'}]x},\\
  |f|^2 &= e^{-2\Imag[k_{e\uparrow,-}]x}+ |r|^2e^{-2\Imag[k_{e\uparrow,+}]x} + 2\Real[re^{i(k_{e\uparrow,+}-k_{e\uparrow,-}^{*})x}],\\ 
  |g|^2 &=|r^{'}|^2 e^{-2\Imag[k_{e\downarrow,+}^{'}]x}.  
\end{aligned}
\end{equation}
Therefore, the probability current is given by
\begin{equation}
\begin{aligned}
    j_P^{\text{AM},e\uparrow} &= (\frac{\hbar k_{e\uparrow,-}}{m_e}+ \frac{\alpha k_y}{\hbar})e^{-2\Imag[k_{e\uparrow,-}]x} \\
    &+ (\frac{\hbar k_{e\uparrow,+}}{m_e}+ \frac{\alpha k_y}{\hbar})|r|^2e^{-2\Imag[k_{e\uparrow,+}]x} + \frac{\hbar}{m_e}\Real[(k_{e\uparrow,+}+k_{e\uparrow,-}^{*})re^{i(k_{e\uparrow,+}-k_{e\uparrow,-}^{*})x}] +\frac{2\alpha k_y}{\hbar}\Real[re^{i(k_{e\uparrow,+}-k_{e\uparrow,-}^{*})x}]\\&+ (\frac{\hbar k_{e\downarrow,+}^{'}}{m_e}- \frac{\alpha k_y}{\hbar})|r^{'}|^2 e^{-2\Imag[k_{e\downarrow,+}^{'}]x}.
\end{aligned}
\label{eq:jP_AM}
\end{equation}

On the other hand, the probability current in the FI is given by
\begin{equation}
    j_P^{\text{FI}} = \frac{\hbar}{m_e} [\Imag{\{f^*\nabla f \}} + \Imag{\{g^*\nabla g \}}].
\end{equation}

For $\Psi_{\text{FI},e\uparrow}$, we have $f = t a_+ e^{-ik_{\text{F}1,e\uparrow}x} + p a_- e^{-ik_{\text{F}2,e\uparrow}x} $ and $g = t b_+ e^{-ik_{\text{F}1,e\uparrow}x} + p b_- e^{-ik_{\text{F}2,e\uparrow}x}$. Since the wavevectors in the FI are imaginary, we apply $k_{\text{F}1,e\uparrow}=i\kappa_1$ and $k_{\text{F}2,e\uparrow}=i\kappa_2$ where $\kappa_1$ and $\kappa_2$ are real. Consequently, we have $f = t a_+ e^{\kappa_1x} + p a_- e^{\kappa_2x} $ and $g = t b_+ e^{\kappa_1 x} + p b_- e^{\kappa_2x}$.
\begin{equation}
  \Imag{\{f^* \nabla f\}}= \Imag{\{\kappa_1 |t|^2|a_+|^2 e^{2\kappa_1x} + \kappa_2 |p|^2|a_-|^2 e^{2\kappa_2x} + (\kappa_1 a_+ a_{-}^{*} tp^{*} + \kappa_2 a_{+}^{*}a_-t^{*}p)e^{(\kappa_1+\kappa_2)x}\}}.
\label{eq:f*f approx}
\end{equation}
It is obvious that the first two terms in Eq. (\ref{eq:f*f approx}) are zero since $\kappa_1$ and $\kappa_2$ are real. If $\kappa_1 = \kappa_2$, we can have $\Imag{\{f^* \nabla f\}}=0$. However, we should have $\kappa_1 \neq \kappa_2$ according to the exchange $J$ in Eqs. (\ref{eq:kF1_eup},\ref{eq:kF2_eup}) of FI. Therefore, we have   
\begin{equation}
  \Imag{\{f^* \nabla f\}}= \Imag{\{(\kappa_1 a_+ a_{-}^{*} tp^{*} + \kappa_2 a_{+}^{*}a_-t^{*}p)e^{(\kappa_1+\kappa_2)x}\}}.
\end{equation}
Similarly, we have
\begin{equation}
  \Imag{\{g^* \nabla g\}}= \Imag{\{(\kappa_1 b_+ b_{-}^{*} tp^{*} + \kappa_2 b_{+}^{*}b_-t^{*}p)e^{(\kappa_1+\kappa_2)x}\}}.
\end{equation}
Consequently, the probability current is given by
\begin{equation}
j_P^{\text{FI},e\uparrow} = \frac{\hbar}{m_e}\Imag{\{[\kappa_1 (a_+ a_{-}^{*}+b_+ b_{-}^{*}) tp^{*} + \kappa_2 (a_{+}^{*}a_-+b_{+}^{*}b_-)t^{*}p]e^{(\kappa_1+\kappa_2)x}\}}.
\label{eq:jP_SC}
\end{equation}
Note here there are no separate terms regarding the transmission coefficients $t$ and $p$ but the mixing terms between them.

Apply $j_P^{\text{FI},e\uparrow}  = j_P^{\text{AM},e\uparrow} $ and insert Eqs. (\ref{eq:jP_AM},\ref{eq:jP_SC}),
\begin{equation}
\begin{aligned}
    1 &= A(E) + B(E) + C(E),\\
    A(E) &= -\frac{(\frac{\hbar k_{e\uparrow,+}}{m_e}+ \frac{\alpha k_y}{\hbar})|r|^2e^{-2\Imag[k_{e\uparrow,+}]x} + \frac{\hbar}{m_e}\Real[(k_{e\uparrow,+}+k_{e\uparrow,-}^{*})re^{i(k_{e\uparrow,+}-k_{e\uparrow,-}^{*})x}] +\frac{2\alpha k_y}{\hbar}\Real[re^{i(k_{e\uparrow,+}-k_{e\uparrow,-}^{*})x}]}{(\frac{\hbar k_{e\uparrow,-}}{m_e}+ \frac{\alpha k_y}{\hbar})e^{-2\Imag[k_{e\uparrow,-}]x}},\\
    B(E) &= -\frac{(\frac{\hbar k_{e\downarrow,+}^{'}}{m_e}- \frac{\alpha k_y}{\hbar})|r^{'}|^2 e^{-2\Imag[k_{e\downarrow,+}^{'}]x}}{(\frac{\hbar k_{e\uparrow,-}}{m_e}+ \frac{\alpha k_y}{\hbar})e^{-2\Imag[k_{e\uparrow,-}]x}},\\
    C(E) &= \frac{\frac{\hbar}{m_e}\Imag{\{[\kappa_1 (a_+ a_{-}^{*}+b_+ b_{-}^{*}) tp^{*} + \kappa_2 (a_{+}^{*}a_-+b_{+}^{*}b_-)t^{*}p]e^{(\kappa_1+\kappa_2)x}\}}}{(\frac{\hbar k_{e\uparrow,-}}{m_e}+ \frac{\alpha k_y}{\hbar})e^{-2\Imag[k_{e\uparrow,-}]x}}=0,  
\end{aligned}   
\end{equation}
in which $A(E)$, $B(E)$ and $C(E)$ are the probability coefficients regarding reflection without spin-flip ($r$), reflection with spin-flip ($r^{'}$) and transmission ($t$ and $p$), respectively. Note that $C(E)$ becomes zero when Eq. (\ref{eq:ab}) is employed. Next, we will focus on the spin-flip probability regarding $B(E)$, which plays an important role in spin pumping.

\begin{figure}[t!]
\includegraphics[width=0.99\columnwidth]{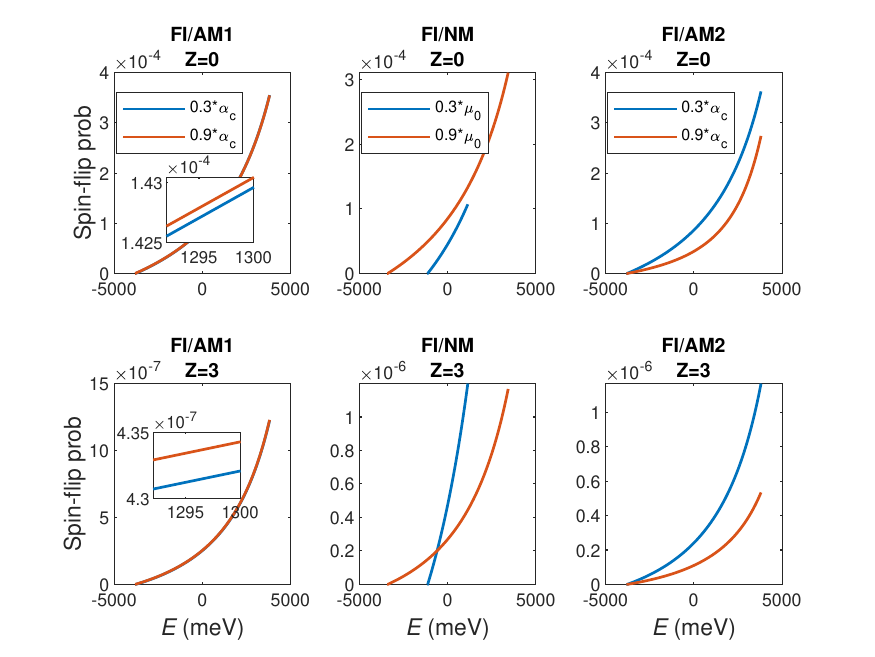}
	\caption{(Color online) Spin-flip probability for different spin pumping bilayers for $Z=0$ and $Z=3$ at a fixed $k_y$ mode. Here a small $k_y=0.1$ is utilized.}
	\label{fig:prob}
\end{figure}

In Fig. \ref{fig:prob}, the spin-flip probability is plotted for different spin pumping bilayers including FI/AM1(AM2) for $Z=0$ and $Z=3$, in which it is found that the spin-flip probability increases (decreases) with altermagnetism in FI/AM1(AM2). On the other hand, it is shown that the spin-flip probability decreases with $\mu$ in FI/NM for large $Z$. These observations are consistent with the spin pumping current behavior shown in Fig. 2 in the main text.

\section{Arbitrary-angle rotated AM}
The arbitrary-angle rotated AM can be modeled based on the combination of our established AM1 and AM2 cases, i.e., a more general Hamiltonian is
\begin{equation}
    {H}_\text{AM} =  -{\frac{\hbar^2\triangledown^2}{2 m_e}} - \mu + \alpha_1 k_x k_y{\sigma_z} + \alpha_2 (k_x^2- k_y^2){\sigma_z}/2,
\end{equation}
in which two different altermagnetism strength parameters $\alpha_1$ and $\alpha_2$ are introduced and the arbitrary angle is determined by $\theta_\alpha = \frac{1}{2}\arctan(\alpha_1/\alpha_2)$. Following the same procedure as introduced before, the eigenvalues and wave vectors can be solved from the Hamiltonian, e.g.,
\begin{equation}
E_\pm=\frac{\hbar^2(k_{x}^2+k_{y}^2)}{2m_e}-\mu \pm \alpha_1 k_xk_y\pm \frac{\alpha_2}{2} (k_x^2-k_y^2),
\end{equation}
\begin{align}
    k_{e\uparrow,\pm}=&\pm\frac{1}{\hbar+
    \alpha_2m_e/\hbar}\sqrt{2m_e(\mu+E)(1+\frac{\alpha_2m_e}{\hbar^2})-\hbar^2 k_y^2 +\frac{(\alpha_1^2+\alpha_2^2) m_e^2 k_y^2}{\hbar^2}}\notag\\
    &-\frac{\alpha_1 m_e k_y}{\hbar^2+m_e\alpha_2},
\end{align}
which reveal features of both the AM1 and AM2 cases. To ensure that the energy dispersion corresponds to an elliptical energy surface rather than a hyperbola, the altermagnetism parameters should satisfy $\bar \alpha \equiv \sqrt{\alpha_1^2+\alpha_2^2} < \alpha_c \equiv \hbar^2/m_e$. The corresponding semi-major and semi-minor axes are $a=\sqrt{\frac{2m_e(\mu+E)}{\hbar^2-m_e\bar\alpha }} $ and $b=\sqrt{\frac{2m_e(\mu+E)}{\hbar^2+m_e\bar\alpha }} $ for electron incidents, based on which the DOS can be calculated. Similarly, the boundary conditions and spin current expressions can be derived from the Hamiltonian with all necessary details included in our previous explanation for the AM1 and AM2 cases.

\end{widetext}

\end{document}